\documentclass[twocolumn]{aastex631}%,linenumbers

\newcommand{\haopt}{\Delta _\mathrm{H\alpha-opt}}
\newcommand{\uvopt}{\Delta _\mathrm{UV-opt}}
\newcommand{\uvoptonly}{\Delta _\mathrm{UV-opt, noH\alpha}}

\newcommand{\rr}[1]{#1}

\newcommand \msun{\mathrm{M}_{\odot}}

\newcommand \rcorecorr{r_{\mathrm{core,corr}}}

\usepackage{comment}
\usepackage{booktabs}
\usepackage{hhline}
\usepackage{multirow}
\usepackage[shortlabels]{enumitem}

\shorttitle{AASTeX v6.3.1 Sample article}
\shortauthors{GASP team}

\begin{document}
\def\linefilter{F$680$N$_{\mathrm{line}}$\,}

\title{Morphology of star-forming clumps in ram-pressure stripped galaxies as seen by HST}

\author[0000-0002-3818-1746]{Eric Giunchi}
\affiliation{INAF-Osservatorio Astronomico di Padova, Vicolo Osservatorio 5, 35122 Padova, Italy}
\affiliation{Dipartimento di Fisica e Astronomia, Universit\`a di Padova, Vicolo Osservatorio 3, 35122 Padova, Italy}

\author[0000-0001-8751-8360]{Bianca M. Poggianti}
\affiliation{INAF-Osservatorio Astronomico di Padova, Vicolo Osservatorio 5, 35122 Padova, Italy}

\author[0000-0002-7296-9780]{Marco Gullieuszik}
\affiliation{INAF-Osservatorio Astronomico di Padova, Vicolo Osservatorio 5, 35122 Padova, Italy}

\author[0000-0002-1688-482X]{Alessia Moretti}
\affiliation{INAF-Osservatorio Astronomico di Padova, Vicolo Osservatorio 5, 35122 Padova, Italy}

\author[0000-0002-4382-8081]{Ariel Werle}
\affiliation{INAF-Osservatorio Astronomico di Padova, Vicolo Osservatorio 5, 35122 Padova, Italy}

\author[0000-0001-8600-7008]{Anita Zanella}
\affiliation{INAF-Osservatorio Astronomico di Padova, Vicolo Osservatorio 5, 35122 Padova, Italy}

\author[0000-0003-0980-1499]{Benedetta Vulcani}
\affiliation{INAF-Osservatorio Astronomico di Padova, Vicolo Osservatorio 5, 35122 Padova, Italy}

\author[0000-0002-8710-9206]{Stephanie Tonnesen}
\affiliation{Flatiron Institute, CCA, 162 5th Avenue, New York, NY 10010, USA}

\author[0000-0002-5189-8004]{Daniela Calzetti}
\affiliation{Department of Astronomy, University of Massachusetts, 710 N. Pleasant Street, LGRT 619J, Amherst, MA 01002, USA}

\author[0000-0002-6179-8007]{Callum Bellhouse}
\affiliation{University of Nottingham School of Physics and Astronomy, University Park, Nottingham, NG7 2RD}
\affiliation{INAF-Osservatorio Astronomico di Padova, Vicolo Osservatorio 5, 35122 Padova, Italy}

\author[0000-0002-9136-8876]{Claudia Scarlata}
\affiliation{Minnesota Institute for Astrophysics, School of Physics and Astronomy, University of Minnesota, 316 Church Street SE, Minneapolis, MN 55455, USA}

\author[0000-0002-8372-3428]{Cecilia Bacchini}
\affiliation{INAF-Osservatorio Astronomico di Padova, Vicolo Osservatorio 5, 35122 Padova, Italy}

\begin{abstract}
We \rr{characterize} the morphological properties of a \rr{statistically relevant sample} of H$\alpha$ and UV young star-forming clumps and optical complexes, observed with the \textit{Hubble Space Telescope} in six galaxies of the GASP sample undergoing ram-pressure stripping.
The catalogs comprise 2406 (323 in the tails) H$\alpha$ clumps, 3750 (899) UV clumps and 424 tail optical complexes. About 15-20\% of the clumps and 50\% of the complexes are resolved in size.
\rr{We find that} more than half of the complexes contain no H$\alpha$ clumps, while most of them contain at least one UV clump. The clump number and size increase with the complex size\rr{, while} the median complex filling factor is larger for UV clumps ($0.27$) than for H$\alpha$ clumps ($0.10$) and does not correlate with almost any morphological property. This suggests that the clumps number and size grow with the complex keeping the filling factor constant.
\rr{When studying the position of the clumps inside their complexes, }H$\alpha$ clumps, and UV clumps to a lesser extent, show a displacement from the complex center of $0.1-1$ kpc and, in $\sim 60$\% of the cases, they are displaced away from the galactic disk. This is in accordance with the fireball configuration, already observed in the tails of stripped galaxies.
Finally, the filling factor and the clump radius increase with the distance from the galactic disk, suggesting that the reciprocal displacement of the different stellar generations increases as a consequence of the velocity gradient caused by ram pressure.
\end{abstract}


\section{Introduction}\label{sec:intro}
Galaxies in clusters undergo a large variety of environmental processes that strongly influence their evolution.
The higher density of galaxies with respect to the field increases the probability of high-speed close encounters, eventually triggering a process called harassment in which the tidal interaction heats the stellar kinematics and in some cases strips the gas out of the galactic disk \citep{Moore1996,Moore1998}.
Other processes include starvation/strangulation, with the gas in the galactic halo removed by the interaction with the intracluster medium (ICM; \citealt{Larson1980,Balogh2000}) and ram-pressure stripping (RPS; \citealt{Gunn1972}), in which the pressure exerted by the ICM is so strong to strip the interstellar medium (ISM), forming streams (tails) of gas as long as hundreds of kiloparsecs. The galaxies showing this peculiar morphology are often called \textit{jellyfish} galaxies.

In particular, RPS is shown to eventually quench the star formation in the galaxy, as a consequence of the removal of the gas reservoir \citep{Vulcani2020a,Boselli2022review}. However, observations and simulations have shown that RPS can trigger both a short burst of star formation in the galactic disks (\citealt{Vulcani2018,Vulcani2019,Vulcani2020} for observational data, \citealt{Goller2023} for evidence from TNG simulation), most likely as a consequence of the gas compression, and in-situ formation of new stars in the tails of stripped ISM, occurring in compact clumps (\citealt{Smith2010,Merluzzi2013,Abramson2014,Kenney2015,Consolandi2017,Jachym2019, Poggianti2019a} for observations, \citealt{Tonnesen2010,Tonnesen2012,Tonnesen2019} for simulations).

The presence of newborn stars in the tails of these galaxies represents a unique laboratory to study the star formation in a regime in which the molecular gas is out of the galactic plane and embedded in the hot, low-metallicity and high-pressure ICM, shedding light on the mechanism responsible for the collapse of the molecular gas into stars.

One of the most peculiar structures found in these star forming regions are the so-called \textit{fireballs} \citep{Cortese2007,Yoshida2012,Kenney2014,Jachym2017,Waldron2023}, in which the different stages of star formation are spatially displaced from each other, from the pre-collapse molecular gas clumps, to actively star forming clumps, to stellar-only clumps not actively forming new stars \citep{Poggianti2019a}. These stages are spatially correlated in coherent elongated structures a few hundreds of parsecs long, in which the early ones are typically displaced further from the galactic disk than the late ones.
What is thought to occur is that, as the stars form in the stripped gas, they do not feel the ram pressure exerted by ICM anymore, being a collisionless system. Therefore they retain their initial velocity, the one that the stripped gas had at the moment of the star-formation episode.
On the other hand, the ICM keeps decelerating the stripped gas, introducing a difference in velocity between gas and stars.
As a consequence of that, the stellar and gaseous component decouple and once the stripped gas undergoes new star-formation events, the young generations of stars that are formed are located further from the galactic disk than the old ones. The result is that the stellar component is spatially distributed in elongated structures and the stellar age anti-correlates with the distance from the galaxy \citep{Kenney2014}.

A systematic study of how the star formation occurs in these clumps and which are the differences from those observed in the disk of star-forming galaxies, main-sequence galaxies is therefore necessary.
This is one of the aims of the GASP project (GAs Stripping Phenomena in galaxies with MUSE, \citealt{Poggianti2017a}), a MUSE ESO Large Program that observed galaxies affected by different processes, including gas removal, from the field, through groups, to clusters. This includes 114 galaxies, 64 of which are observed at different RPS stages, from pre-stripping to post-stripping \citep{Fritz2017}. Targets were chosen from the catalog in \cite{Poggianti2016} as galaxies with long unilateral tail-like structures in B-band images.
The galaxies of the final sample have masses in the range $10^9-10^{11.5}\,\msun$ and redshift between 0.04 and 0.07. 
MUSE allowed a spatially resolved study of the properties of the ionized gas phase and the stellar component both in the galactic disks and in the stripped tails. 

One of the main results of GASP is the detection of ongoing star formation in the tails of a large sample of jellyfish galaxies, as probed by the presence of strong H$\alpha$ emission ionized by very young stars \citep{Bellhouse2017,Gullieuszik2017,Moretti2018b,Moretti2020,Poggianti2019b}.
However, the morphology of these star-forming clumps could not be resolved due to the limited spatial resolution of the MUSE observations ($\sim1$ kpc).

In order to improve the spatial resolution with which these clumps are observed, six galaxies of the GASP sample have been observed with the UVIS channel of the WFC3 onboard \textit{HST} \citep{Gullieuszik2023}, whose PSF angular size is $\sim0.07\arcsec$ ($\sim14$ times better than the seeing limited GASP MUSE observations). The broad-band filters which were adopted cover a spectral range going from UV- to I-band restframe, and are the F275W, F336W, F606W and F814W. The targets were observed also with the narrow-band filter F680N, from which it is possible to extract the H$\alpha$ emission.
Details about the clump detection and selection are given in \cite{Giunchi2023} and summarized in Sec. \ref{sec:dataset}.

This multi-wavelength dataset allows us to \rr{perform the first statistically relevant study of} the morphology of stellar clumps of different ages, from H$\alpha$ emitting clumps tracing ongoing star formation to those that have stopped forming new stars, observable in U- and V-bands. The goal of this paper is to \rr{contribute to our understanding of} how the different stellar generations are correlated and nested within each other, and fully characterize the properties of the stellar populations in the tails of these jellyfish galaxies\rr{, by exploiting our large sample of clumps}.
Finally, thanks to the exquisite angular resolution of HST images, we can test whether these stellar populations are arranged according to the fireball model previously described.
\rr{A full characterization of the morphology of these objects and of the correlations between the properties of clumps and complexes at different wavelengths is necessary to infer the future of these objects, whether they will remain bound to the parent galaxy or be lost in the cluster and whether they will evolve as compact structures or fade away becoming more and more diffuse. The fate of the complexes is beyond the scope of this paper, and will be studied in future works.}

This paper is structured as follows. In Sec. \ref{sec:dataset} the \textit{HST} data and the clump catalogs are presented. In Sec. \ref{sec:morphology} we define the morphological properties that are studied. Sec. \ref{sec:results} is focused on the axial ratio distribution of the clumps and on how clumps detected at different wavelength are nested within each other. In Sec. \ref{sec:fireballs} we investigate the presence of fireballs in the tails of the targets. In Sec. \ref{sec:radial} we search for trends of the studied properties with the distance from the parent galaxy. Finally, in Sec. \ref{sec:conclusions}, we summarize our results.

This work adopts standard cosmology parameters $H_0=70\,\mathrm{km\,s^{-1}\,Mpc^{-1}}$, $\Omega_M=0.3$ and $\Omega_\Lambda=0.7$. All the coordinates are reported in the J2000 epoch. The strength of the correlations is evaluated following the guidelines by \cite{Evans1996}.

\section{Dataset}\label{sec:dataset}
The six targets were selected for Hubble Space Telescope (\textit{HST}) observations \citep{Gullieuszik2023} for their particularly long H$\alpha$ emitting tails (between 40 and 120 kpc) and the large number of H$\alpha$ clumps ($\sim 250$) detected with MUSE observations (\citealt{Poggianti2019a}, see also \citealt{Vulcani2020}). The properties of our sample of galaxies are summarized in Tab. \ref{tab:targets}.

\begin{deluxetable*}{cccccccccc}
\tablecaption{Main properties of our target galaxies and of their clusters.}
\tablewidth{0pt}
\tablehead{
\colhead{ID$_{P16}$} & \colhead{RA} & \colhead{Dec} & \colhead{$M_*$} & \colhead{$R_e$} & \colhead{$z_{gal}$} & \colhead{cluster} & \colhead{$z_{clus}$} & \colhead{$\sigma_{clus}$} & \colhead{ref}\\
& \colhead{(J2000)} & \colhead{(J2000)} & \colhead{$\mathrm{10^{10}\,M_{\odot}}$} & kpc &&&&$\mathrm{km/s}$&
}
\decimalcolnumbers
\startdata
JO$175$ & 20:51:17.593 & -52:49:22.34 & $3.2\pm0.5$ & $3.24^{+0.37}_{-0.34}$ & 0.0468 & A3716 & 0.0457 & $753^{+36}_{-38}$ & (4,10,14)\\
\multirow{2}{*}{JO$201$} & \multirow{2}{*}{00:41:30.295} & \multirow{2}{*}{-09:15:45.98} & \multirow{2}{*}{$6.2\pm0.8$} & \multirow{2}{*}{$7.73^{+0.60}_{-0.79}$} & \multirow{2}{*}{0.0446} & \multirow{2}{*}{A85} & \multirow{2}{*}{0.0559} & \multirow{2}{*}{$859^{+42}_{-44}$} & (3,4,5,6,8,9,10,\\ 
&&&&&&&&& 13,14,15,16,18)\\
JO$204$ & 10:13:46.842 & -00:54:51.27 & $4.1\pm0.6$ & $4.58^{+0.22}_{-0.26}$ & 0.0424 & A957 & 0.0451 & $631^{+43}_{-40}$ & (2,4,6,12,18)\\ 
JO$206$ & 21:13:47.410 & +02:28:35.50 & $9.1\pm0.9$ & $8.94^{+1.08}_{-1.14}$ & 0.0513 & IIZW108 & 0.0486 & $575^{+33}_{-31}$ & (1,4,6,10,13,18,19)\\ 
JW$39$ & 13:04:07.719 & +19:12:38.41 & $17\pm3$ & $10.16^{+1.23}_{-1.34}$ & 0.0650 & A1668 & 0.0634 & 654 & (14,18,19,21)\\
\multirow{2}{*}{JW$100$} & \multirow{2}{*}{23:36:25.054} & \multirow{2}{*}{+21:09:02.64} & \multirow{2}{*}{$29\pm7$} & \multirow{2}{*}{$7.35^{+0.43}_{-0.74}$} & \multirow{2}{*}{0.0602} & \multirow{2}{*}{A2626} & \multirow{2}{*}{0.0548} & \multirow{2}{*}{$650^{+53}_{-49}$} & (4,6,7,10,11,\\
&&&&&&&&& 15,17,18,19,20)\\
\enddata
\tablecomments{Columns are: GASP ID of the galaxy as in \citealt{Poggianti2016} (1), RA and Dec of the galaxy (2 and 3), galaxy stellar mass (4), effective radius (5), galaxy redshift (6), ID of the host cluster (7), cluster redshift (8), cluster velocity dispersion (9), references (10). References are: 1) \cite{Poggianti2017a}, 2) \cite{Gullieuszik2017}, 3) \cite{Bellhouse2017}, 4) \cite{Poggianti2017b}, 5) \cite{George2018}, 6) \cite{Moretti2018b}, 7) \cite{Poggianti2019b}, 8) \cite{Bellhouse2019}, 9) \cite{George2019}, 10) \cite{Radovich2019}, 11) \cite{Moretti2020}, 12) \cite{Deb2020}, 13) \cite{Ramatsoku2020}, 14) \cite{Bellhouse2021}, 15) \cite{Tomicic2021a}, 16) \cite{Campitiello2021}, 17) \cite{Ignesti2022a}, 18) \cite{Tomicic2021b}, 19) \cite{Ignesti2022b}, 20) \cite{Sun2021}, 21) \cite{Peluso2022}. Masses are taken from \cite{Vulcani2018}. Effective radii are taken from \cite{Franchetto2020} and converted in kpc. Cluster redshifts and velocity dispersions are taken from \cite{Biviano2017} and \cite{Cava2009}.}
\label{tab:targets}
\end{deluxetable*}

The observations were performed using the WFC3/UVIS on board of \textit{HST} in 4 broad-band photometric filters (F275W, F336W, F606W, F814W) and one narrow-band filter, F680N. The broad-band filters guarantee a spectral coverage from UV- to I-band restframe, while the narrow-band filter collects the H$\alpha$ emission line at the redshift of the targets (Tab. \ref{tab:targets}).
The pixel angular size is $0.04\arcsec$. 
The UVIS PSF is almost constant in all the filters, with a FWHM of $\sim 0.07\arcsec$\footnote{https://hst-docs.stsci.edu/wfc3ihb/chapter-6-uvis-imaging-with-wfc3/6-6-uvis-optical-performance}. Such FWHM corresponds to $\sim 70\,\mathrm{pc}$ at the redshifts of the clusters hosting the targets.
Further details about exposure times, sensitivity of the images, data reduction and analysis are given in \cite{Gullieuszik2023}, as well as a precise description of the steps adopted to extract the H$\alpha$ emission from the F680N.

\subsection{Distinction of disk, extraplanar and tail regions}\label{sec:spatcat}
For each galaxy, clumps born from the stripped gas (\textit{tail} region) are studied separately from those formed in the galaxy (\textit{disk} region).
Furthermore, a third spatial category, \textit{extraplanar}, is distinguish the clumps that are located in the disk but have elongated shape suggestive of an alignment along a preferential direction. These elongated clumps are likely affected by ram pressure but fall in the disk region because of projection effects.
Further details are given in \cite{Giunchi2023}.

\subsection{Catalog of star-forming clumps and complexes}\label{sec:clumps}

In this section we summarize the properties of the samples of star-forming clumps and complexes.
For a more complete description, see \cite{Giunchi2023}, in which we also studied the properties luminosity and size distribution function and their luminosity-size relation.

The two samples of star-forming clumps are detected and selected independently in H$\alpha$-line and UV (F275W filter, $\sim 260$ nm rest-frame at z$=0.05$) images. Both these tracers probe recently formed stars, tracing the line emission of the gas ionized by massive OB stars and stellar continuum from OBA stars, respectively. H$\alpha$ emission traces star formation on timescales $\sim 10$ Myr, while UV emission on timescales $\sim 200$ Myr \citep{Kennicutt1998a,Kennicutt2012,Haydon2020}, allowing us to study how star formation in clumps changes both in time and space.
Clumps are selected using the Python package \textsc{Astrodendro}\footnote{https://dendrograms.readthedocs.io/en/stable/index.html} so that each clump candidate has at least 5 contiguous pixels with flux above 2$\sigma$\footnote{$\sigma$ is the single-pixel detection limit, computed as described in \cite{Gullieuszik2023}.} in the detection band.
Moreover, \textsc{Astrodendro} can build a hierarchical tree structure by merging bright sub-structures (leaves) in large structures (trunk).
From the initial sample detected by \textsc{Astrodendro}, only candidates with a signal-to-noise ratio (SNR) larger than 2 in at least 3 out of 5 photometric bands are kept.
We used MUSE observations to confirm detections in the tails and exclude background objects (see \citealt{Poggianti2019a}). Candidates outside of the MUSE field of view are confirmed if they have SNR$>2$ in all the 5 photometric bands and positive H$\alpha$ emission.

The final samples include 2406 H$\alpha$-selected clumps (1708 in the disks, 375 extraplanar, 323 in the tails) and 3745 UV-selected clumps (2021 in the disks, 825 extraplanar, 899 in the tails). Only about $2\%$ of the clumps are trunks containing leaves.

The visual inspection of tails H$\alpha$- and UV-selected clumps pointed out that many UV and H$\alpha$ clumps were connected in large structures observed in optical wavelengths, which we call \textit{star-forming complexes}.
In order to detect the complexes, which are likely to be populated also by stars older than those found in H$\alpha$- and UV-selected clumps, we ran \textsc{Astrodendro} on the F606W optical broad-band ($\sim 560$ nm rest-frame at z$=0.05$). The detection is performed only in the galaxy tails and increasing the flux threshold from 2$\sigma$ to 3$\sigma$ in order to better separate the single complexes.
Only trunk candidates are kept, since the aim is probing the whole stellar content of structures formed from stripped gas. Furthermore, each complex candidate must be matched\footnote{As described in \cite{Giunchi2023}, a complex is \textit{matched} to a clump if their areas overlap by at least one pixel.} with at least one H$\alpha$- or UV-selected clump in order to be confirmed. These selection criteria result in a sample of 424 star-forming complexes.

\section{Morphological quantities of clumps and complexes}\label{sec:morphology}

As detailed in \cite{Giunchi2023}, a variety of physical parameters can be derived for each clump in the sample (or catalogue). Fluxes are computed by integrating within the whole area of the clump/complex, and sizes are defined as twice the PSF-corrected core radius ($\rcorecorr$). The core radius is the geometric mean of the semi-major and semi-minor axes of the clumps, computed as the standard deviation of the clump surface brightness distribution along the direction of maximum elongation and the perpendicular direction, respectively. These axes are also used to define the position angle of the clumps as the angle between the RA-axis and the major axis of the clump.

To obtain a robust estimate of some quantities, like the size of the clumps or the axial ratio, the clump/complex must be resolved.
A clump or a complex is classified as resolved if its size is larger than $2\mathrm{FWHM_{PSF}}$ ($\sim 140$ pc at the redshift of our galaxy sample). The H$\alpha$- and UV-resolved clumps are typically $15-20\%$ of the whole sample, depending on the detection filter and the spatial category (i.e. tail, disk, extraplanar), while about half of the star-forming complexes (204) are resolved. Thus, in the following we will use 286 H$\alpha$-resolved (224 disk, 37 extraplanar and 25 tail clumps), 591 UV-resolved clumps (369 disk, 118 extraplanar, 104 tail clumps).

For each clump and complex, in this paper we also compute the following:

\begin{enumerate}
    \item distance D from the center of the galaxy: since the tail clumps are formed from the gas stripped from the galactic disk, it is not possible to infer their position along the line of sight and therefore to de-project their position. For this reason, this is the projected distance in the plane of the sky. 
    \item axial ratio AR: ratio between the semi-minor and semi-major axes of the clump. According to this definition, elongated clumps have a smaller axial ratio than round ones;
    \item tilt angle $\Delta\theta$: angle between the major axis of the clump/complex and the line connecting the center of the clumps to that of the galaxy.
    The angle is defined between 0 (semi-major axis aligned with the center of the galaxy) and 90 (semi-major axis perpendicular to the direction to the center of the galaxy) degrees. In Appendix \ref{app:subtail}, the same quantity is computed as the angle between the major axis of the clump/complex and the local direction of the sub-tail the clump/complex belong to. Both definitions give consistent results;
    \item F606W luminosity $\mathrm{L_{F606W}}$ integrated inside the mask of the clump/complex. In order to have a reliable estimate of $\mathrm{L_{F606W}}$ for clumps, we select only H$\alpha$- and UV-selected clumps with F606W signal-to-noise larger than 2. Since star-forming complexes are detected in the F606W, their $\mathrm{L_{F606W}}$ estimate is always reliable;
    \item $\mathrm{H\alpha/UV}$ ratio: computed as $\mathrm{L_{H\alpha}/L_{F275W}}$ for all the clumps and complexes for which $\mathrm{L_{H\alpha}}>0$ and the F275W signal-to-noise is larger than 2.
\end{enumerate}

Furthermore, we calculate for each complex the following quantities based on the properties of the associated clumps:

\begin{enumerate}
    \item $\mathrm{N_{H\alpha}}$ and $\mathrm{N_{UV}}$: number of H$\alpha$- and UV-selected clumps hosted in each complex;
    \item $f_\mathrm{A(H\alpha)}$ and $f_\mathrm{A(UV)}$: H$\alpha$ and UV filling factor of the complexes. It quantifies the fraction of total area of the complex covered by the matched H$\alpha$- and UV-selected clumps (if any). It is computed only for resolved complexes as

    \begin{equation}
    f_\mathrm{A}(\mathrm{j})=\frac{\mathrm{\sum\limits_{i=1}^{\mathrm{N_j}} \mathrm{A_{j,i}}}}{\mathrm{A_{compl}}}
    \end{equation}

    where "j" refers either to H$\alpha$ or UV. The filling factor is computed considering both resolved and unresolved matched clumps therefore the resulting value (if the complex contains at least one unresolved clump) is an upper limit on the intrinsic filling factor, as unresolved star-forming clumps could be smaller than the data resolution;
    \item $r_{bc}(\mathrm{H}\alpha)$ and $r_{bc}(\mathrm{UV})$: respectively, the PSF-corrected core radius of the brightest H$\alpha$- and UV-resolved clump matched to a resolved complex.
\end{enumerate}

For every quantity, Table \ref{tab:nclumps} summarizes the number of H$\alpha$- and UV-selected clumps and optical complexes for which it is possible to derive the geometrical and luminosity properties described above.

\begin{table*}
\centering
\caption{Number of tail H$\alpha$ clumps, UV clumps and star-forming complexes for which each property that is used in this work can be computed. Definitions of the quantities are given in Sec. \ref{sec:morphology}.}
\begin{tabular}{ccc|ccc}
\toprule\toprule
Property & \rr{Symbol} & Condition & H$\alpha$ & UV & Compl.\\\hline
\rr{Clump-galaxy distance} & D & None & 323 & 899 & 424\\
\rr{Clump/complex exact area} & $\mathrm{A_{exact}}$ & Resolved & 25 & 104 & 204\\
\rr{Axial ratio} & AR & Resolved & 25 & 104 & 204\\
\rr{Tilt angle} & $\Delta\theta$ & Resolved & 25 & 104 & 204\\
\rr{F606W luminosity} & $\mathrm{L_{F606W}}$ & $\mathrm{SNR_{F606W}}>2$ & 291 & 878 & 424\\
&&&&&\\
\multirow{2}{*}{\rr{H$\alpha$-to-UV flux ratio}} & \multirow{2}{*}{$\mathrm{H\alpha/UV}$} & $\mathrm{L_{H\alpha}}>0$ \& & \multirow{2}{*}{248} & \multirow{2}{*}{693} & \multirow{2}{*}{327}\\
& & $\mathrm{SNR_{F275W}}>2$ & & & \\\hline
\rr{Number of matched H$\alpha$ clumps} & $\mathrm{N_{H\alpha}}$ & None & - & - & 424\\
\rr{Number of matched UV clumps} & $\mathrm{N_{UV}}$ & None & - & - & 424\\
\rr{H$\alpha$ clumps filling factor} & $f_\mathrm{A}(\mathrm{H}\alpha)$ & Resolved \& $\mathrm{N_{H\alpha}}>0$ & - & - & 108\\
\rr{UV clumps filling factor} & $f_\mathrm{A}(\mathrm{UV})$ & Resolved \& $\mathrm{N_{UV}}>0$ & - & - & 192\\
&&&&&\\
\multirow{2}{*}{\rr{Brightest resolved H$\alpha$ clump}} & \multirow{2}{*}{$r_{bc}(\mathrm{H}\alpha)$} & $\mathrm{N_{H\alpha}}>0$ \& & \multirow{2}{*}{-} & \multirow{2}{*}{-} & \multirow{2}{*}{20}\\
& & one resolved clump & & & \\
&&&&&\\
\multirow{2}{*}{\rr{Brightest resolved UV clump}} & \multirow{2}{*}{$r_{bc}(\mathrm{UV})$} & $\mathrm{N_{UV}}>0$ \& & \multirow{2}{*}{-} & \multirow{2}{*}{-} & \multirow{2}{*}{69}\\
& & one resolved clump & & & \\
\bottomrule
\end{tabular}
\label{tab:nclumps}
\end{table*}

\section{Results}\label{sec:results}
In this section we present the results related to the intrinsic morphology of clumps and complexes. In Sec. \ref{sec:ar}, we explore the axial ratio of clumps and complexes, while in Sec. \ref{sec:nesting} we \rr{study} the nesting of the complexes, studying how the properties of the clumps are correlated to those of the complex hosting them.

\subsection{Axial ratios of clumps and complexes}\label{sec:ar}
The axial ratio of clumps and complexes can be a proxy of the effects of ram-pressure stripping on the gas that is forming stars.
We analyze the axial ratio for both H$\alpha$- and UV-resolved leaf clumps, as well as resolved star-forming complexes.

\begin{figure*}
\includegraphics[width=0.33\textwidth]{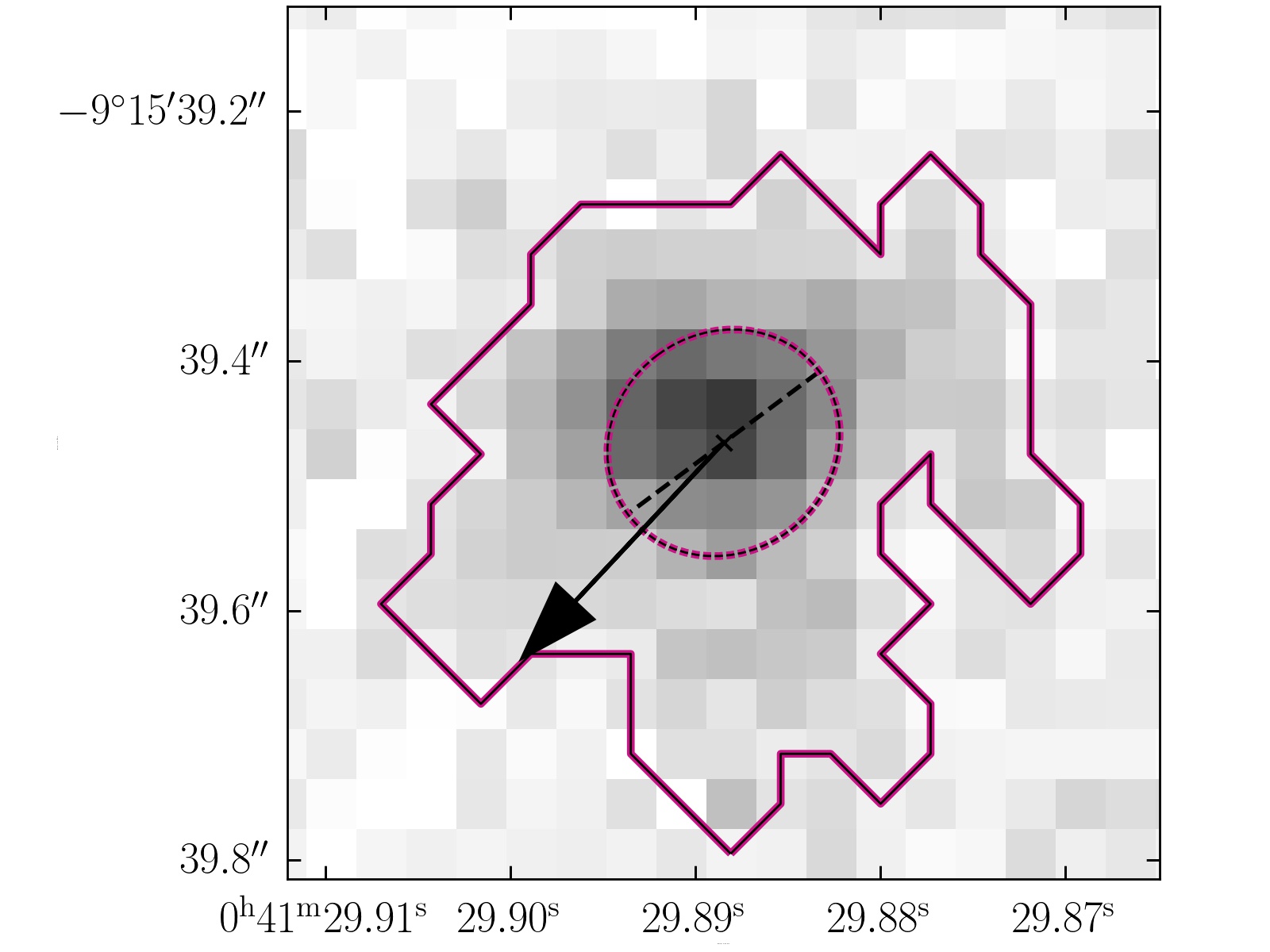}\hfill
\includegraphics[width=0.33\textwidth]{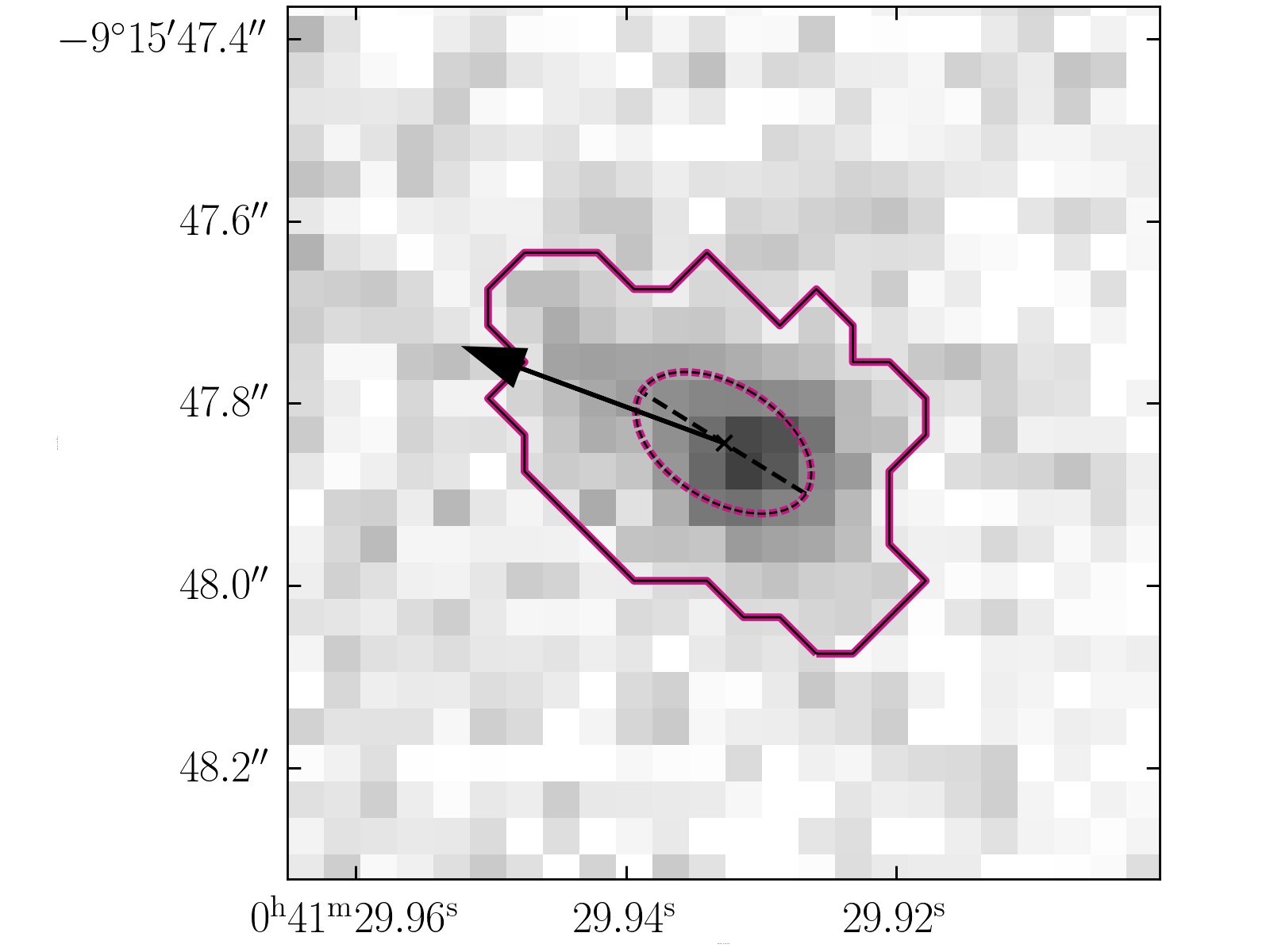}\hfill
\includegraphics[width=0.33\textwidth]{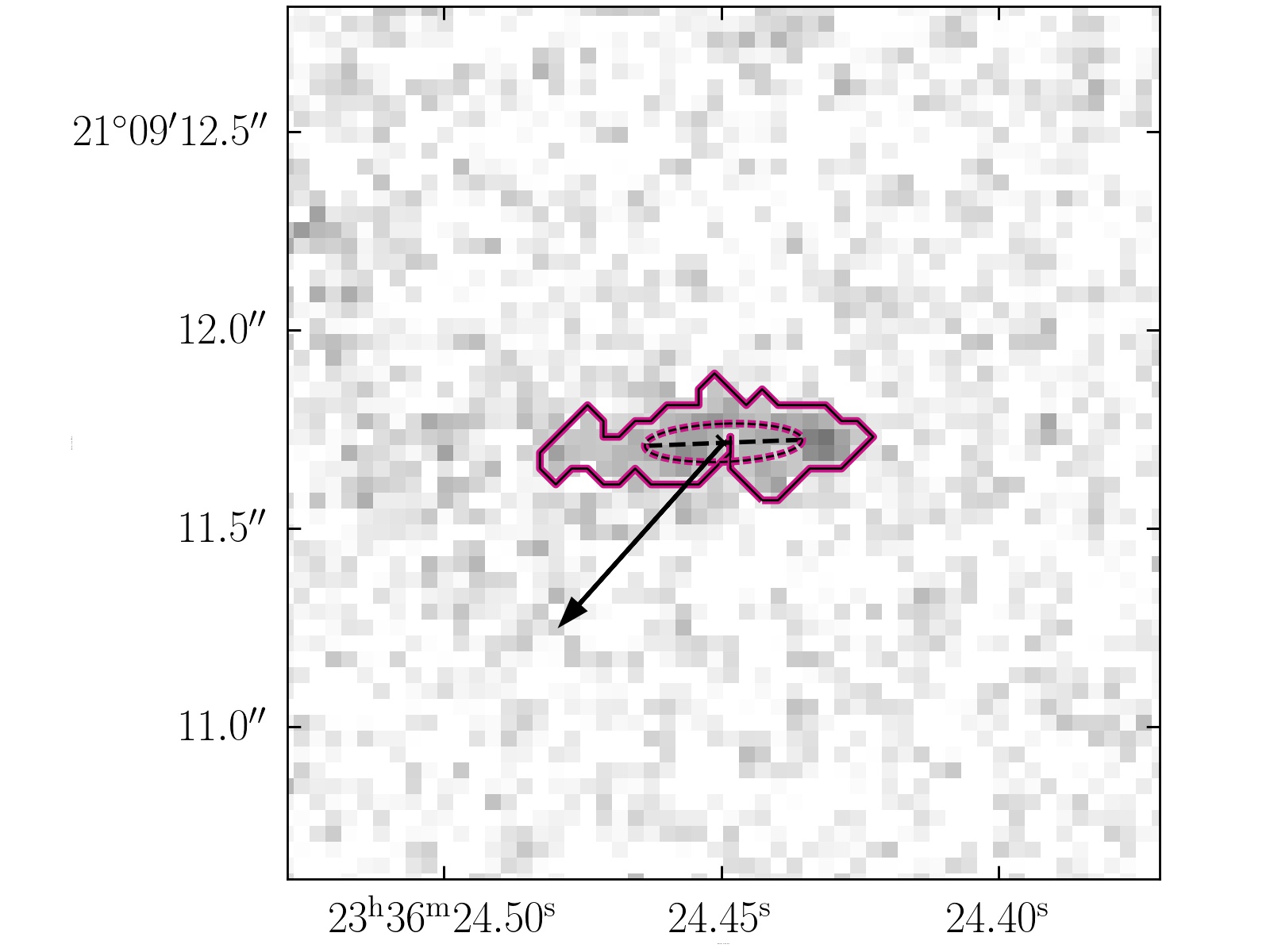}
\caption{Images of 3 UV-resolved clumps. The solid contour defines the area of the clump, the dashed ellipse is defined by the major and minor axis. The dashed line is the major axis. The cross is the geometric center of the clump. The arrow points to the center of the hosting galaxy. The clumps are selected as examples of the variety of elongation we have: from left to right, the axial ratio (defined as the ratio of the semi-minor to the semi-major axes of the clump) decreases from 0.92, to 0.60, to 0.24.
}
\label{plot:clumpelongation}
\end{figure*}

As examples, we show in Fig. \ref{plot:clumpelongation} three UV-resolved clumps with increasing elongation (axial ratios 0.92, 0.60 and 0.24). The solid contour defines the area of the clump, while the dotted ellipse is defined from the semi-major and semi-minor axes of the clump, which also shows the position angle of the clump. The arrow points towards the center of the galaxy, in order to have a qualitative evaluation of the tilt angle.

\begin{figure*}
\includegraphics[width=0.95\textwidth]{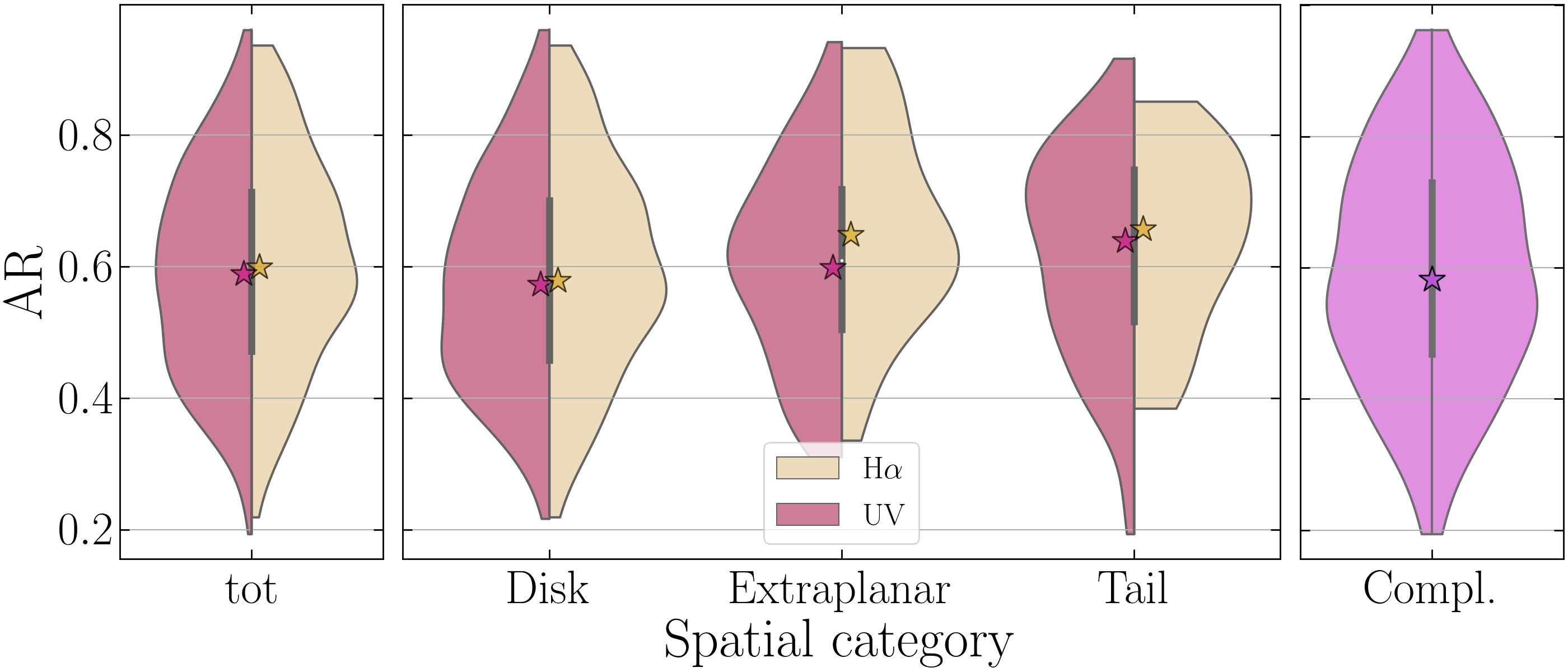}
\caption{Violin plots of the axial ratio distributions of the star-forming clumps and complexes of our sample. Median values are plotted as stars, the dark vertical line shows the interquartile region. Left plot: distribution of the whole UV- (magenta) and H$\alpha$- (golden) resolved clumps. Middle plot: UV- (magenta) and H$\alpha$- (golden) resolved clumps are divided in spatial categories (disk, extraplanar and tail from left to right). Right plot: distribution of the resolved star-forming complexes.}
\label{plot:ar}
\end{figure*}

In Fig. \ref{plot:ar} we show the violin plots of the axial ratio distributions of the H$\alpha$- and UV-resolved leaf clumps and of the resolved complexes.
Median values are in the range $\sim 0.58-0.68$ for all the categories of clumps and complexes.

We also perform a Kolmogorov-Smirnov test by separating H$\alpha$- from UV-selected clumps and complexes, in each of the three spatial categories (tail, extraplanar and disk). We do not find any difference among the sub-sample distributions of the axial ratio, with the only exception of UV-resolved disk and tail clumps.

Nonetheless, we can still derive some conclusions by looking at the distributions in Fig. \ref{plot:ar}. The median values and the peaks of the distributions indicate that clumps tend to be rounder going from the disk to the tails. 
Furthermore, comparing disk and extraplanar clumps, the distribution becomes narrower, suggesting that there are less elongated clumps in the extraplanar regions than in the disk. Indeed, the peak shifts towards rounder clumps (especially in H$\alpha$). In the tails the trend of H$\alpha$-resolved clumps is even narrower and peaking at larger values of AR. The opposite seems happening for tail UV-resolved clumps, which are elongated ($3.33\%$ of the whole tail UV-resolved sample with AR$<0.3$).

In conclusion, we do not see any evidence that tail or extraplanar clumps are more elongated than clumps in the disk. In fact, they are round on average, except for a small fraction of UV-resolved clumps in the tails. However, we point out that also the star-forming complexes, located only in the tails by construction, can reach very low axial ratios. The optical emission comes from stars of different ages, from very young stars emitting in UV and even in H$\alpha$ to stars older than 200 Myr that do not emit neither in UV nor in H$\alpha$. Therefore the cases of strong elongation observed in the complexes is consistent with the fireball model, in which stars of different populations have also different velocity and subsequently increase more and more the reciprocal distance.

\subsection{Multi-wavelength nesting of star-forming complexes in the tails}\label{sec:nesting}

In this section we study the nesting properties of the star-forming complexes, by looking at the number of H$\alpha$- and UV-selected clumps they contain. We stress again that complexes are defined only in the tails, so all the results in this section refer to that spatial category. Clumps are matched to complexes as summarized in Sec. \ref{sec:clumps} and described more in detail in \cite{Giunchi2023}.

\begin{figure*}
\includegraphics[width=0.33\textwidth]{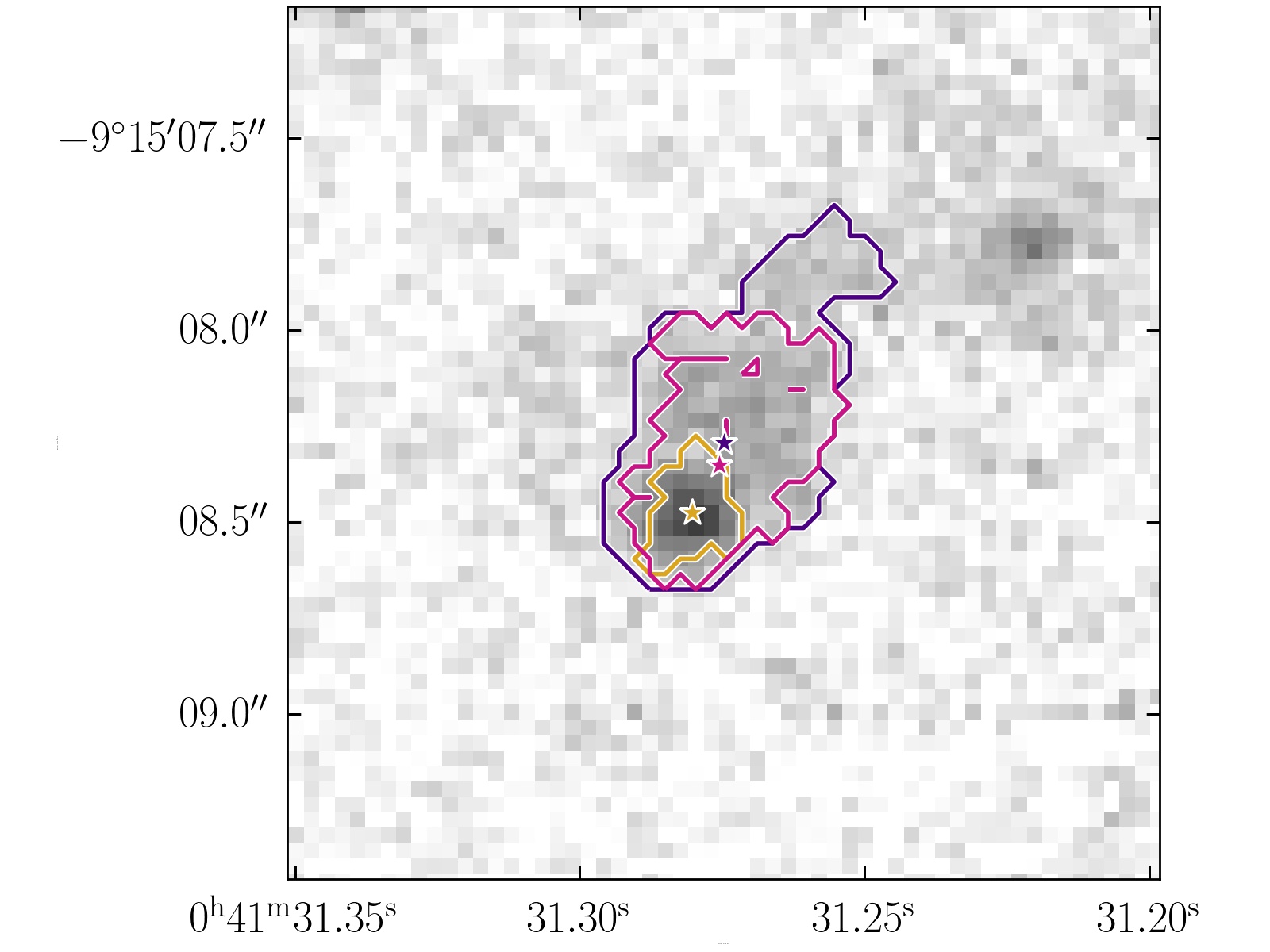}\hfill
\includegraphics[width=0.33\textwidth]{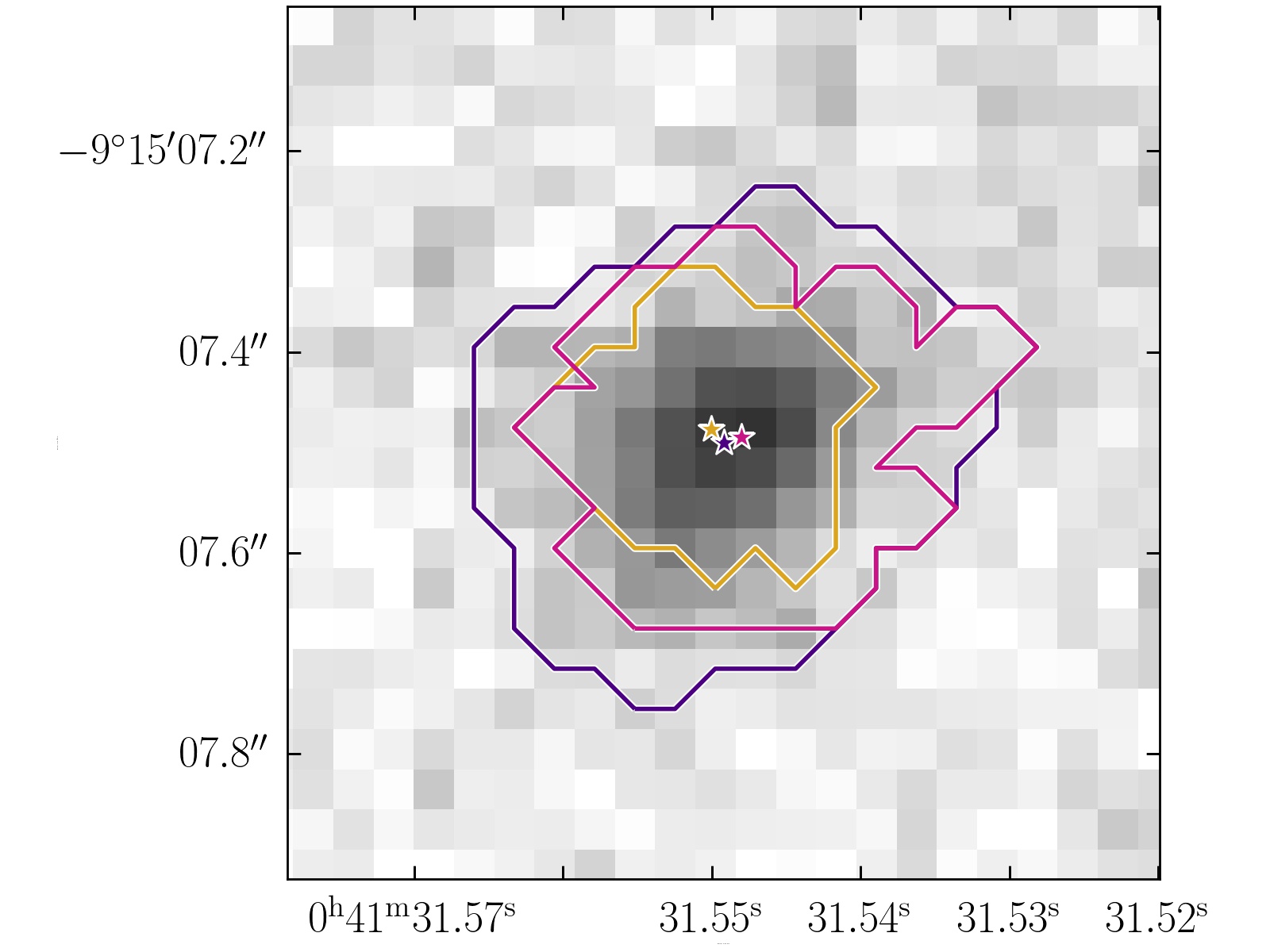}\hfill
\includegraphics[width=0.33\textwidth]{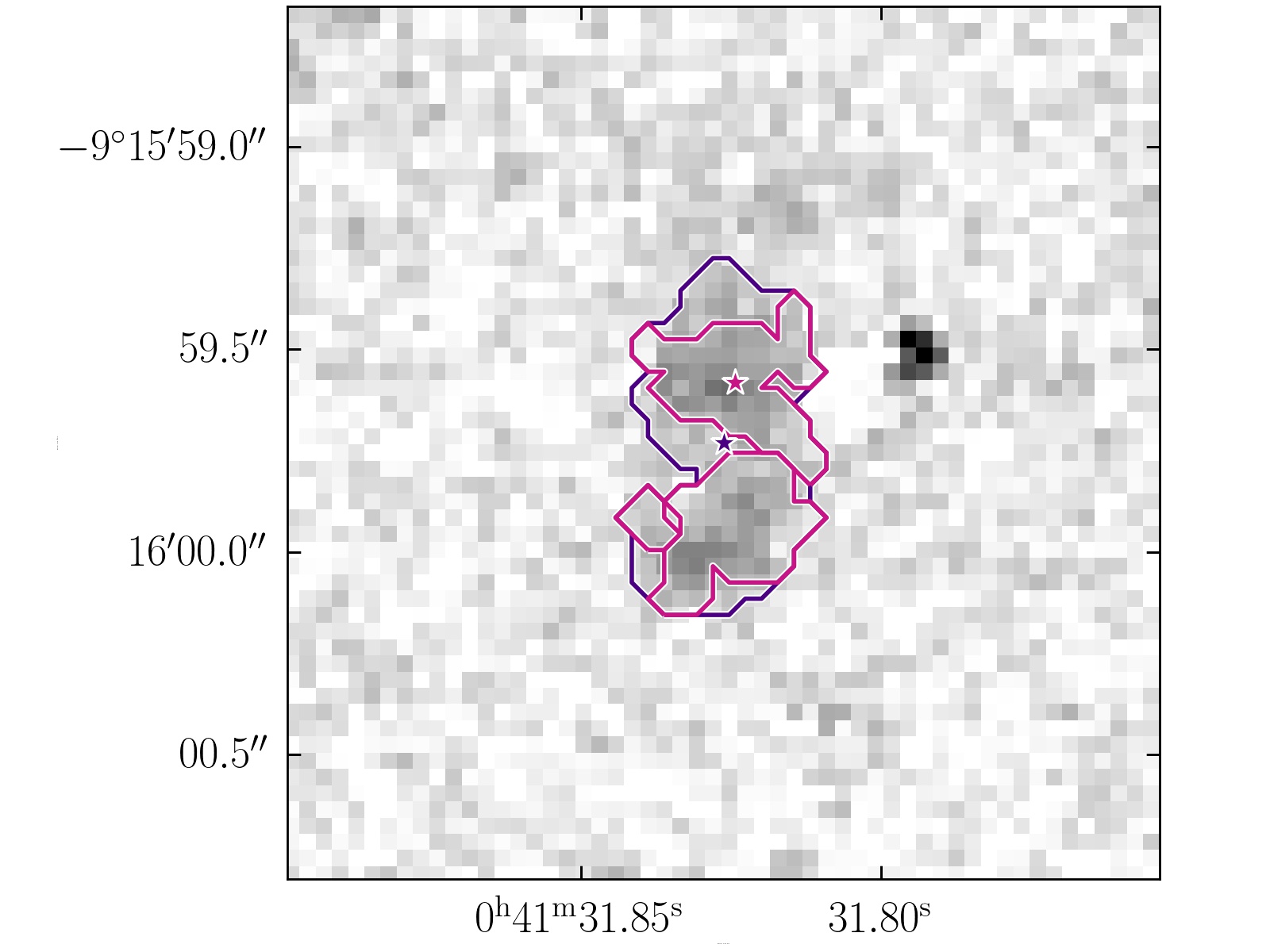}
\caption{Images of 3 star-forming complexes (in dark violet) together with the matched UV- (in purple) and H$\alpha$- (in golden, when present) selected trunk clumps. Geometrical centers of the complex and the brightest H$\alpha$- and UV-selected matched clumps are plotted as violet, purple and golden stars, respectively. Underlying images are in the optical F606W band.
}
\label{plot:clumpnesting}
\end{figure*}

As examples, in Fig. \ref{plot:clumpnesting} three complexes are shown as representatives of the variety of nesting configurations that our sample contains. In the left, we show an elongated complex containing one H$\alpha$-selected clump on one side of it (in gold) and one UV-selected clump (in magenta) covering a larger fraction of the optical emission (in dark violet). In the middle, a similar case but for a round complex is presented. In this case, there is almost no displacement among the centers of the H$\alpha$-selected clump, the UV-selected clump and the optical star-forming complexes. In the right, a complex containing no H$\alpha$-selected clumps and three UV-selected clumps, which cover almost the whole optical complex.

\subsubsection{Number of clumps within complexes}\label{sec:number}

\begin{figure*}
\includegraphics[width=\textwidth]{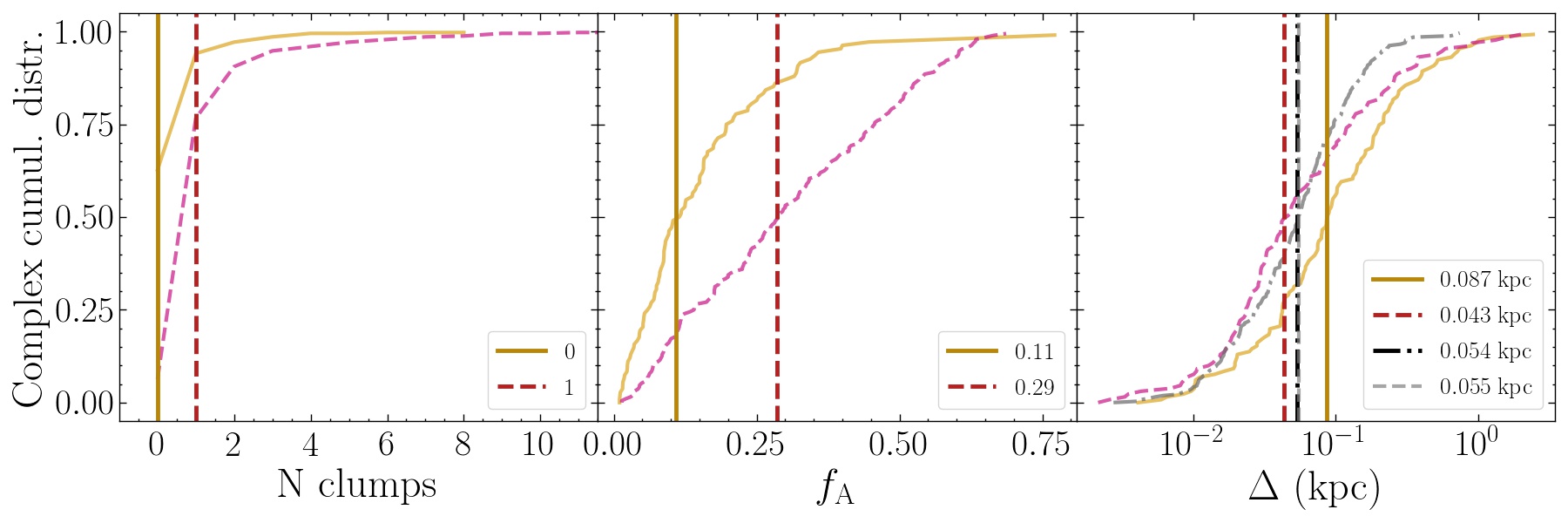}
\caption{Cumulative distributions of the complexes per number of matched clumps (left panel, Sec. \ref{sec:number}), clump filling factor (middle panel, Sec. \ref{sec:fa}) and clump-complex geometric center displacement (right panel, Sec. \ref{sec:displacement}).
Each quantity is computed for matched H$\alpha$-selected clumps (golden solid line) and UV-selected clumps (magenta dashed line). Furthermore, the displacement is computed considering separately UV-selected clumps in complexes with and without matched H$\alpha$ clumps (the latter plotted as grey dotted-dashed line). Median values are plotted as dark golden solid, dark magenta dashed and, when present, black dotted-dashed vertical lines.
In the right panel, the grey dashed line shows our minimum resolvable resolution for the center displacement ($\sim 0.055$ kpc).
}
\label{plot:cumulative}
\end{figure*}

In the top left panel of Fig. \ref{plot:cumulative}
we plot the cumulative distributions of the complexes as a function of the number of hosted H$\alpha$- and UV-selected clumps: 96\% of the complexes contain one or no H$\alpha$-selected clump and 3 UV-selected clumps at most.
It is more common for a complex to host multiple UV-selected clumps rather than multiple H$\alpha$-selected clumps.

The complexes without $\rm H\alpha$ clumps are lacking bright HII regions, though they may still have a detectable diffuse $\rm H\alpha$ emission. Our $\rm H\alpha$-clump detection procedure is sensitive to the emission powered by a single O class-stars
(see Fig. 12 in \citealt{Giunchi2023}).\footnote{An O5V star produces about 5.6$\times 10^{49}$ ionizing photons per second, which for ionization bound conditions corresponds to an $\rm H\alpha$ luminosity of 5.6$\times 10^{37} \rm \, erg \, s^{-1}$.} The lack of O class-stars in these complexes can have two explanations: either star formation has stopped more than 20 Myr ago (all O class-stars have died, the clump is quenched/older), or star formation is still occurring but only stars with masses lower than O or bright B class-stars are currently forming (ongoing star formation with a top-light initial mass function). The presence of faint H$\alpha$ clumps below our detection limit can be another explanation. Nonetheless, in Werle et al. (in prep.), we show by SED-fitting that the complexes devoid of H$\alpha$ clumps are older than the others and than 20 Myr, which makes the presence of undetected H$\alpha$ clumps very unlikely.

\begin{figure*}
\includegraphics[width=\textwidth]{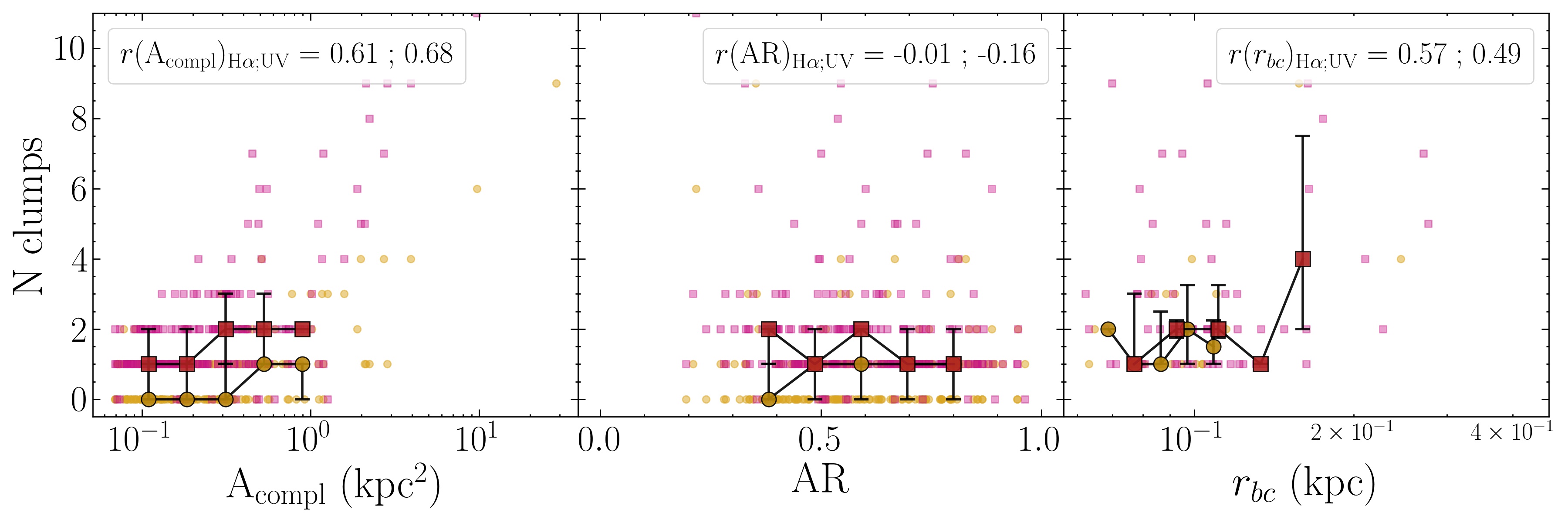}
\caption{\rr{Plots of the number of matched H$\alpha$-selected clumps (golden dots) and UV-selected clumps (magenta squares) as a function of area $\mathrm{A_{compl}}$ (left), axial ratio AR (middle) and radius of the largest matched resolved clump (right).} Larger dark golden dots for H$\alpha$-selected clumps and dark magenta squares for UV-selected clumps show the median number of clumps per bin of the $x-$axis. The bins of the $x-$axis are equally spaced in linear or logarithmic scale according to the physical property. Error-bars show the quartiles of the distribution in the bin. On the top right we show the Pearson coefficient for each quantity and for H$\alpha$ and UV matched clumps. For the sake of the clarity of the plots, we cut out one complex containing 21 clumps, which anyway does not affect our interpretation of the plots.
}
\label{plot:nesting}
\end{figure*}

In Fig. \ref{plot:nesting} we plot the number of matched H$\alpha$- and UV-selected clumps as a function of the following properties of the hosting complex: the complex total area $\mathrm{A_{compl}}$, AR and $r_{bc}$. We also report the Pearson's coefficient $r$ for each quantity and sample of clumps.
The only significant correlations are found with $\mathrm{A_{compl}}$ \rr{and $\mathrm{L_{F606W}}$ (plot not shown)}, for which the Pearson's coefficients are between 0.55 and 0.68 (depending on the chosen compared property and whether computed on the matched H$\alpha$- or UV-selected clumps).

Our analysis thus shows that large and bright complexes host a higher number of clumps, meaning that the clumps contained in large complexes are not necessarily the scaled, large version of those contained in small complexes.
Furthermore, since $\mathrm{L_{F606W}}$ traces also old stellar populations and therefore it is strictly connected to the mass of the complex, the correlation found between this quantity and the number of hosted clumps hints that more massive complexes host a larger number of clumps.
No other correlations are found between the number of hosted clumps and the properties of the complex.

\subsubsection{Filling factor $f_\mathrm{A}$}\label{sec:fa}
We now \rr{focus on} how much of the area of each complex is filled with clumps. In particular, the comparison of UV- and H$\alpha$-selected clumps gives hints about the morphological evolution of stellar populations of different age. As the clumps dynamically evolve, they can either merge or evaporate \citep{Fujii2015}, causing variations in their morphology and in many other properties.
The complex filling factors $f_\mathrm{A}$ computing with respect to H$\alpha$- and UV-selected clumps (Sec. \ref{sec:morphology}) can be used to understand how the different generations of stars are spatially distributed with respect to each other.

In the middle panel of Fig. \ref{plot:cumulative} we plot the cumulative distributions of $f_\mathrm{A}\mathrm{(H\alpha)}$ and $f_\mathrm{A}$(UV).
UV-selected clumps are more likely to cover a larger area than H$\alpha$-selected clumps, 95\% of which have a filling factor smaller than 0.4. Median values are $0.27$ for UV-selected clumps and $0.10$ for H$\alpha$-selected clumps.
This difference suggests that H$\alpha$-selected clumps typically occupy a small and compact region of their parent complexes, while UV-selected clumps are both larger and more numerous, as suggested also by the examples given in Fig. \ref{plot:clumpnesting}.
Furthermore, the UV cumulative distribution grows almost linearly, meaning that the distribution of $f_\mathrm{A}$(UV) is flat up to 0.6.

\begin{figure*}
\includegraphics[width=\textwidth]{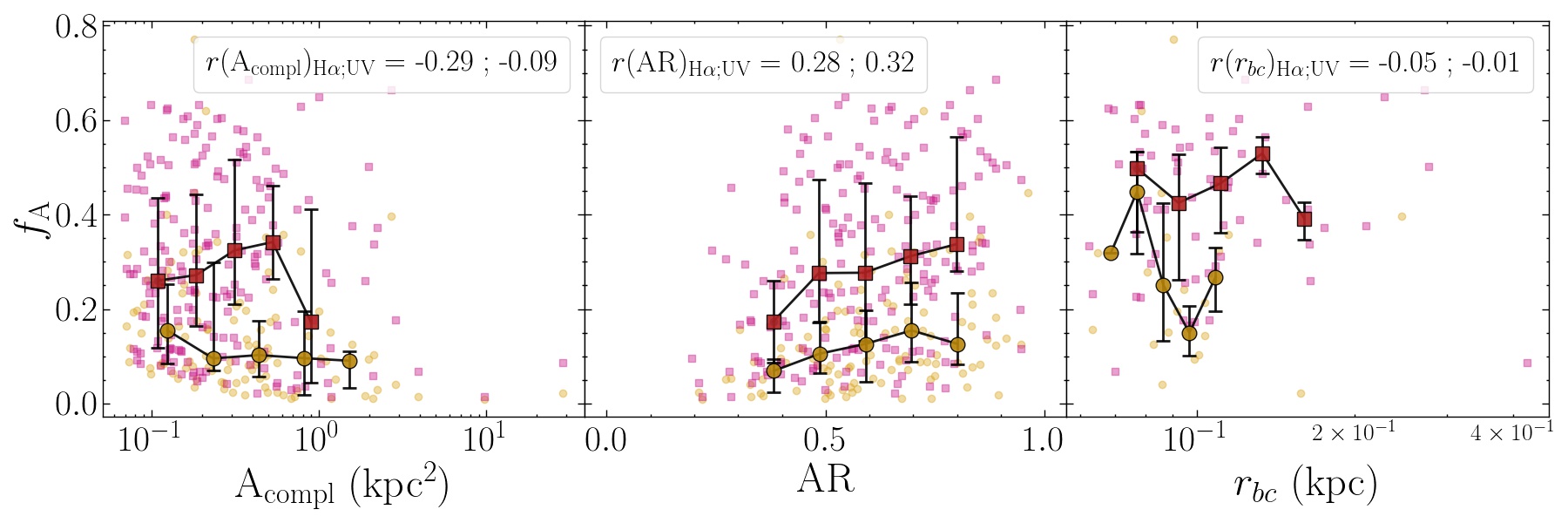}
\caption{As Fig. \ref{plot:nesting}, but for H$\alpha$-selected clumps filling factor $f_\mathrm{A}\mathrm{(H\alpha)}$ (golden dots) and UV-selected clumps filling factor $f_\mathrm{A}\mathrm{(UV)}$ (magenta squares).
}
\label{plot:fA}
\end{figure*}

To further investigate this behaviour, in Fig. \ref{plot:fA} we plot $f_\mathrm{A}$ as a function of $\mathrm{A_{compl}}$, AR and $r_{bc}$.
Weak correlations are found with AR ($r=0.28-0.32$) and $\mathrm{H\alpha/UV}$ ($r=0.42$ for H$\alpha$ clumps\rr{, plot not shown}). No correlations are found with the area or the luminosity of the complex, nor with $r_{bc}$. However, we point out that the spread of the filling factor is large, even where we find a correlation.
Since the filling factor correlates with the axial ratio of the complexes and not with their area, it is likely to be influenced mainly by the intrinsic morphology of the complexes: round complexes, regardless of the size, are better filled by clumps than elongated complexes.

\begin{figure}
\includegraphics[width=0.45\textwidth]{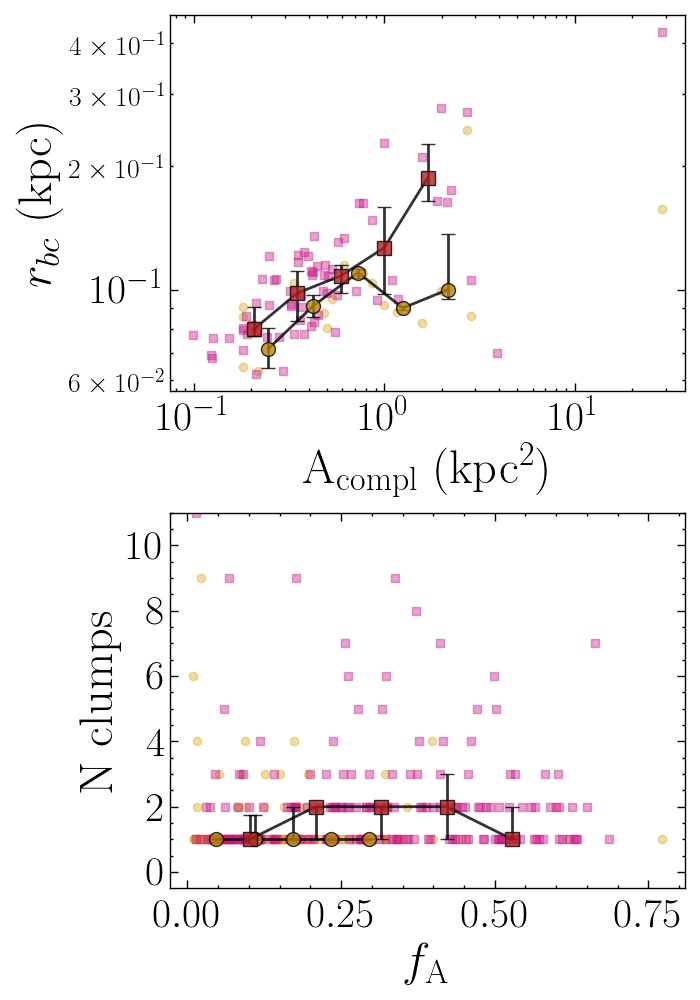}
\caption{Top panel: PSF-corrected core radius of the brightest H$\alpha$-resolved (golden dots) and UV-resolved (magenta squares) clump matched to the complex area. Median profiles are plotted as dark golden dots for H$\alpha$-selected clumps and dark magenta squares for UV-selected clumps. Error-bars show the quartiles of the distribution in the bin.
Bottom panel: same as the top panel, but for number of matched H$\alpha$-selected clumps as a function of the H$\alpha$ filling factor (golden dots) and number of matched UV-selected clumps as a function of the UV filling factor (magenta squares).
}
\label{plot:rcc_nesting}
\end{figure}

In the top panel of Fig. \ref{plot:rcc_nesting} we plot the PSF-corrected core radius of the brightest clump hosted by a complex as a function of the area of the complex itself. A good correlation is observed, both when looking at the H$\alpha$-resolved clumps and at the UV-resolved clumps ($r=0.56$ and $0.76$, respectively). Therefore large complexes both contain more and larger clumps than the small ones\rr{, rather than having more clumps with a constant size or a fixed number of clumps of increasing size}.

However, we find that the filling factor does not correlate with \rr{$\mathrm{A_{compl}}$ and $r_{bc}$ (Fig. \ref{plot:fA}), and neither with the number of clumps, as shown }in the bottom panel of Fig. \ref{plot:rcc_nesting}.
Our conclusion is that, in large complexes, both the size and the number of the nested clumps increase in such a way that the filling factor is not correlated with the area of the complex.
The only morphological quantity driving variations of the filling factor is the axial ratio of the complex: round complexes are more filled by clumps than elongated complexes. \rr{This effect may be due to differences in the RPS deceleration of round and elongated complexes, with the latter being more decelerated and resulting in higher reciprocal velocities among stars born at different times, with a consequent larger displacement among the stellar different generations. Otherwise, the difference may be caused by projection effects: the velocity of elongated complexes shall have the dominant component on the plane of the sky, while the round ones shall preferentially move along the line of sight.}

\section{Fireballs}\label{sec:fireballs}

We now compare our results with the theoretical expectations for fireballs \citep{Kenney2014,Jachym2019}.
The model of the \textit{fireball} predicts that the age of the stellar populations formed in the stripped gas decreases with increasing distance from the galaxy disk.

\begin{figure*}
\begin{center}
\includegraphics[width=0.8\textwidth]{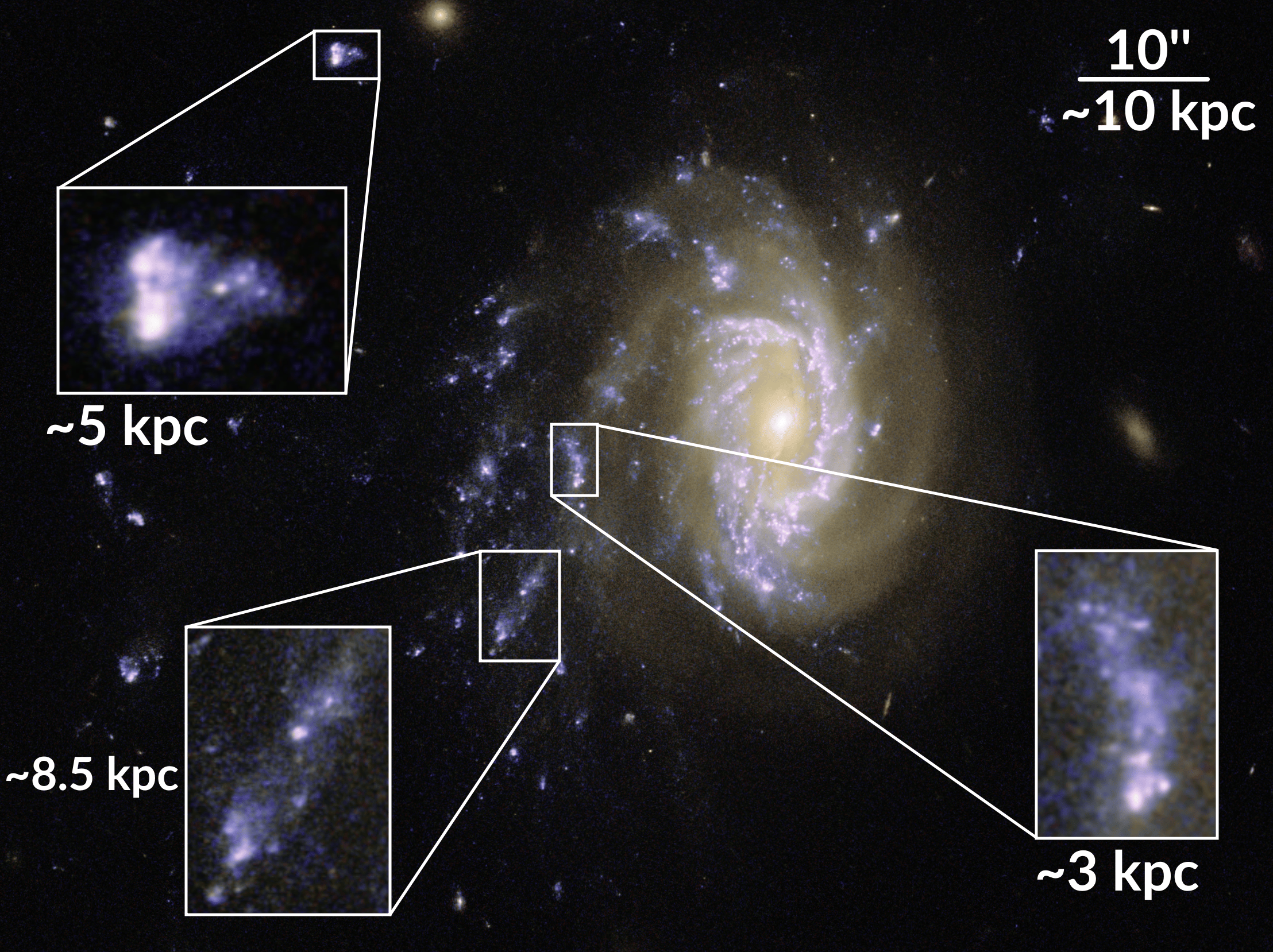}
\end{center}
\gridline{\resizebox{1.03\textwidth}{!}{\includegraphics[height=1cm]{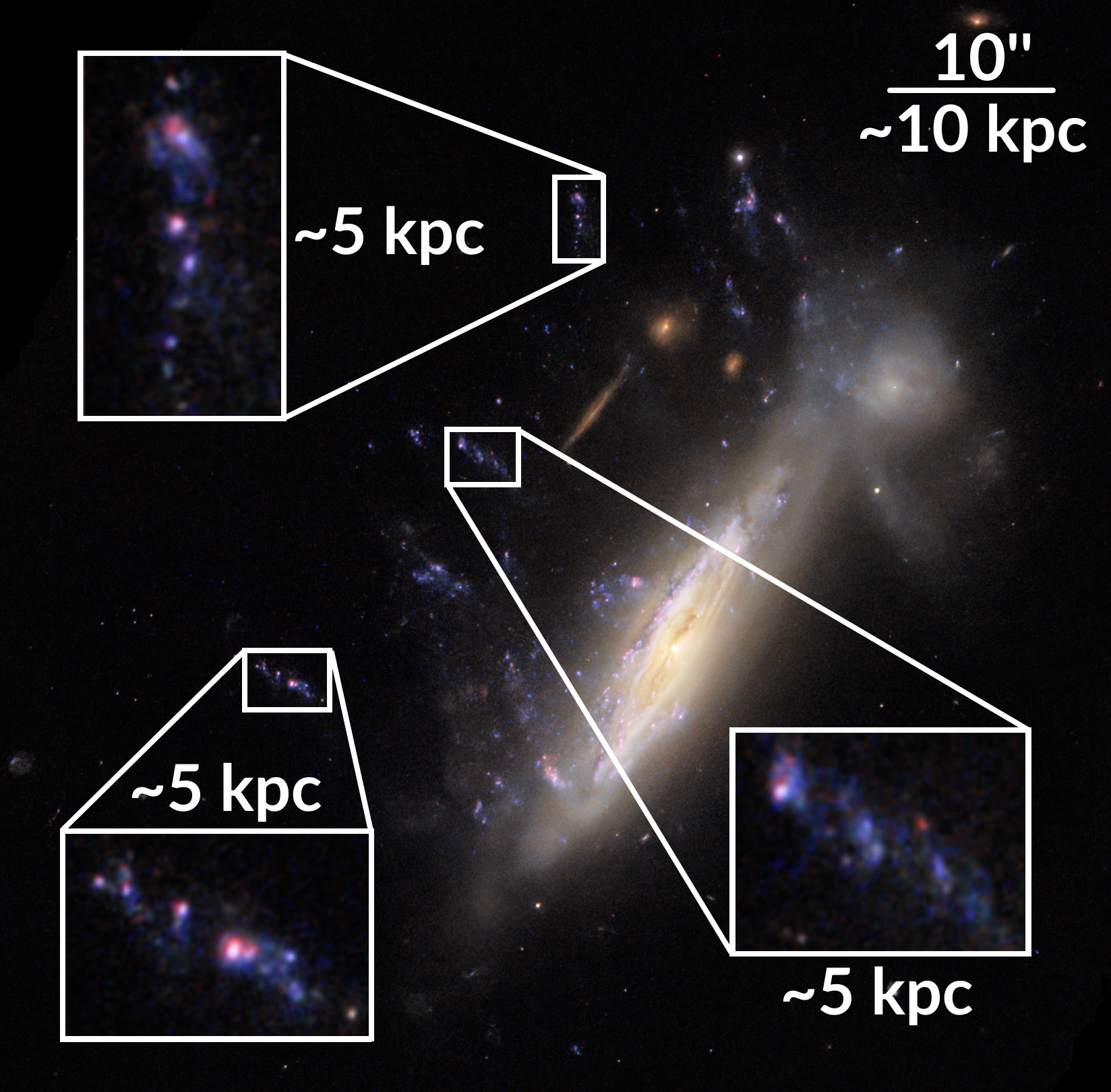}
          \includegraphics[height=1cm]{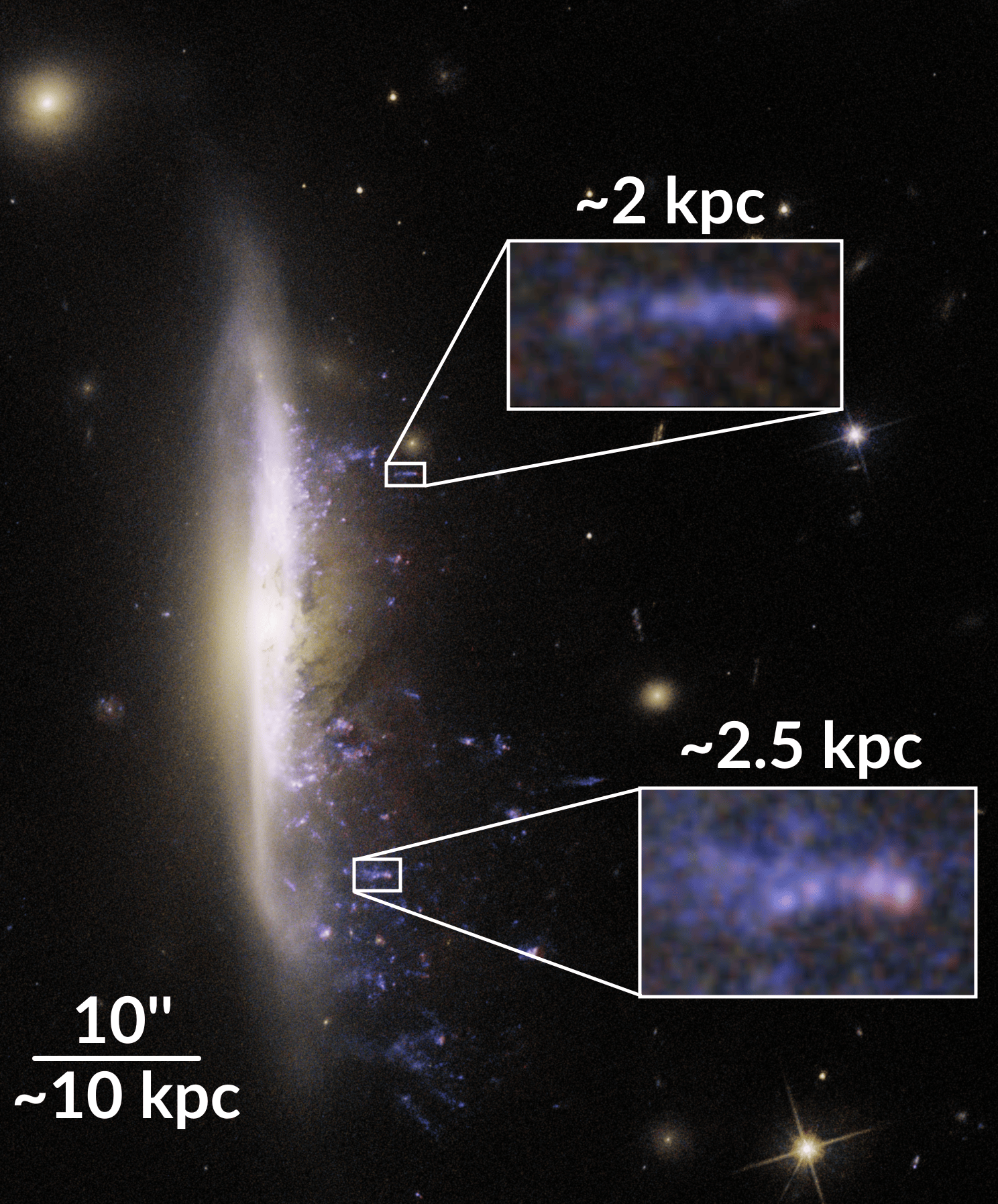}
          }
          }
\caption{Color-composite images of JO201 (top), JO204 (left) and JW100 (right). The orange/yellow color traces the optical emission, the blue one is the UV emission. The red color is associated to H$\alpha$ emission. Zoomed-in images show typical examples of fireball configurations of the star-forming clumps in the tails of these galaxies. Credits for images: ESA/Hubble \& NASA, M. Gullieuszik and the GASP team.}
\label{plot:rgb}
\end{figure*}

In Fig. \ref{plot:rgb} we show eight clear examples of groups of fireballs, three in JO201, three in JO204 and two in JW100, respectively. The fireball associations have size of a few kpc and are all characterized by extended optical and UV emission, with compact bright H$\alpha$ clumps. The H$\alpha$ emission is always displaced in the opposite direction with respect to the center of the galaxy. On the other hand the UV light, despite more compact in correspondence of the H$\alpha$ emission, is more diffuse and characterized by a long tail pointing towards the galaxy center. The optical emission is fainter and more diffuse than the optical emission.
In the case of JO201, the fireballs are elongated either along a direction aligned with the center of the galaxy or along the spiral arm-like sub-tails that are likely to characterize the large-scale morphology of the stripped material (in Appendix \ref{app:subtail} we define the sub-tails for each galaxy and study possible trends along it). In a single fireball many compact H$\alpha$ clumps are found.
The fireballs seen in JO204 appear less diffuse and even more clustered than those observed in JO201, with smaller UV tails but more fragmented star-forming regions. On the other hand, JW100, shows either simple and long fireballs with one H$\alpha$ clump located in the head of the system and a long UV tail (top sub-image), or fireballs with many UV clumps aligned with each other and embedded in a more diffuse UV region.

In the next section, we will quantify the displacement of the H$\alpha$- and UV-selected clumps with respect to the center of their star-forming complexes, to assess the fraction of complexes for which the displacement is oriented as expected by the fireball model, with the H$\alpha$-selected clumps displaced far from the galactic disk.

\subsection{Displacement of clumps inside complexes}\label{sec:displacement}

In agreement with the fireball model, we expect the H$\alpha$ emission to be displaced on one side of the star forming complexes defined by the F606W emission; the UV emission should be de-centered as well, but less than the H$\alpha$ one since it covers a larger age interval (i.e. stars younger than 200 Myr).
To quantitatively estimate this effect, we compute the distance between the geometric centers of the complex and the brightest H$\alpha$-selected clump matched to it ($\haopt$)
and the brightest UV-selected clump ($\uvopt$).
We select only resolved complexes. We have shown in Sec. \ref{sec:morphology} that many complexes do not contain any H$\alpha$-selected clumps, therefore in such cases we compute only $\uvopt$, named $\uvoptonly$ hereafter, and keep these complexes separated from those containing both H$\alpha$- and UV-selected clumps.

The right panel of Fig. \ref{plot:cumulative} shows the cumulative distributions of the displacements $\haopt$, $\uvopt$ and $\uvoptonly$.
The distribution of $\haopt$ favors large displacement and indeed the median value\rr{ ($86.5$ pc)} is the largest among the three samples, while the one of $\uvopt$ can reach very small values, of a few parsecs.\rr{ The median value of $\haopt$ is in good agreement with those observed by \cite{Kenney2014} for H$\alpha$-UV displacements ($80-100$ pc) of 10 clumps in IC3418.}
The UV displacement in complexes with no H$\alpha$-selected clumps, $\uvoptonly$, rises more steeply than the other two distributions and does not reach values larger than $700-800$ pc, even though the median\rr{ displacement ($53.8$ pc)} is larger than the one of $\uvopt$\rr{ ($43.3$ pc)}.
However, almost half of the distances are below the precision on the reciprocal position of the centers of the two components, which we quantified as the sum in quadrature of the size of two pixels ($\sim 0.055$ kpc at z$\simeq 0.05$).

\begin{figure*}
\includegraphics[width=\textwidth]{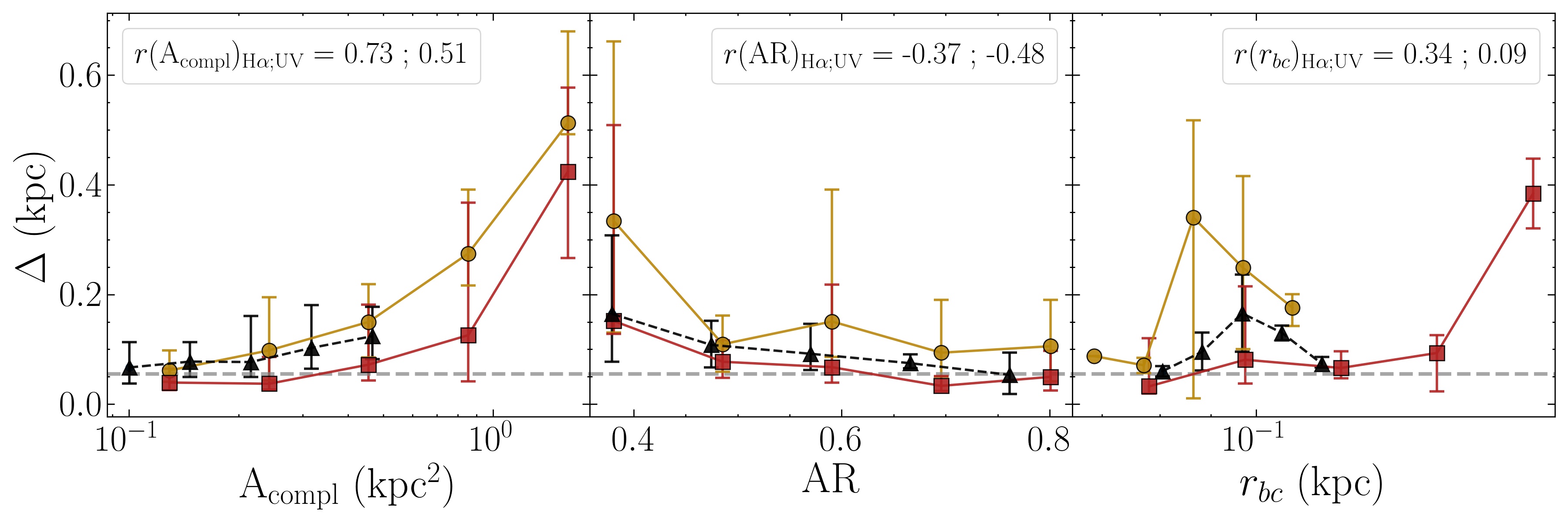}
\caption{\rr{Plots of the center distances of the complexes as a function of area $\mathrm{A_{compl}}$ (left), axial ratio AR (central) and radius of the largest matched resolved clump (right).} We plot the median profiles of the center distances as golden dots for $\haopt$, magenta squares for $\uvopt$, black triangle for $\uvoptonly$. Error-bars show the quartiles of the distribution in the bin. On the top right we show the Pearson coefficient for each quantity and for H$\alpha$ and UV matched clumps. We also plot our resolution limit as a horizontal grey dashed line.}
\label{plot:fireball}
\end{figure*}

In order to understand what drives this trend, in Fig. \ref{plot:fireball} we correlate the reciprocal distances with the usual complex properties. 
From these plots we can conclude that:

\begin{enumerate}
    \item \rr{the best correlations are found with $\mathrm{A_{compl}}$ (Pearson's value $0.73$) and $\mathrm{L_{F606W}}$ ($0.61$, plot not shown).} The larger (and consequently also brighter) the complex, the larger the displacement. In particular, the largest values are reached by $\haopt$, while $\uvopt$ shows a similar but shallower trend and $\uvoptonly$ is flat;
    \item a weak anti-correlation (Pearson's values $\sim -0.38$) is found with AR: the more elongated the complex, the larger the displacement. Also in this case, $\haopt$ shows the steeper correlation, while $\uvopt$ and $\uvoptonly$ have a similar, flatter trend;
    \item no correlation is found with $r_{bc}$(clump), even if here the statistics may be too poor to drive any conclusion.
\end{enumerate}

It is worth mentioning that in the cases of complex area and luminosity, $\uvoptonly$ overlaps remarkably well with $\haopt$, at least in the range of values in which both complexes with and without H$\alpha$ clumps are observed.
Such good agreement may hint that the UV-only complexes have similar morphological properties to those that contains H$\alpha$ clumps, too. As a consequence of passive ageing, the UV emission is now located only in the head of the fireball, where the youngest population resides and where younger complexes emit in H$\alpha$ as well. That may explain why the morphological properties of the UV clumps in UV-only complexes resemble those of H$\alpha$ clumps in the others. To fully answer to this question, a detailed analysis of the ages of these objects is necessary and will be perform in future works.

In conclusion large and bright complexes with both H$\alpha$- and UV-selected clumps show clear signs of a displaced H$\alpha$ emission, where the UV emission is more diffused and closer to the complex center, in accordance with the fireball model previously described.
The displacement decreases with the complex axial ratio, suggesting that in elongated complexes the clumps tend to occupy one side of the complex, instead of being positioned at its center.

\subsection{Alignment of clumps and complexes with the center of the galaxy}\label{sec:tilt}

In accordance with the fireball model, we expect the clumps and the complexes to be aligned along a common direction. For the following analysis, such direction is defined as the one to the center of the hosting galaxy, even if we tested also the directions of the sub-tails defined in Appendix \ref{app:subtail}, finding no substantial differences between the two.

In order to understand whether the clumps and the complexes are elongated along the direction to the galaxy center, we studied two possible proxies: 1) the distributions of the tilt angles of clumps and complexes $\Delta\theta$, defined in Sec. \ref{sec:morphology}; 2) the comparison of projected distances from the center of the galaxy of the H$\alpha$, UV and optical centers of each complex, to understand whether the UV- and especially the H$\alpha$-selected clumps are farther from the disk than the optical emission, as expected by the fireball model.

\begin{figure*}
\includegraphics[width=0.95\textwidth]{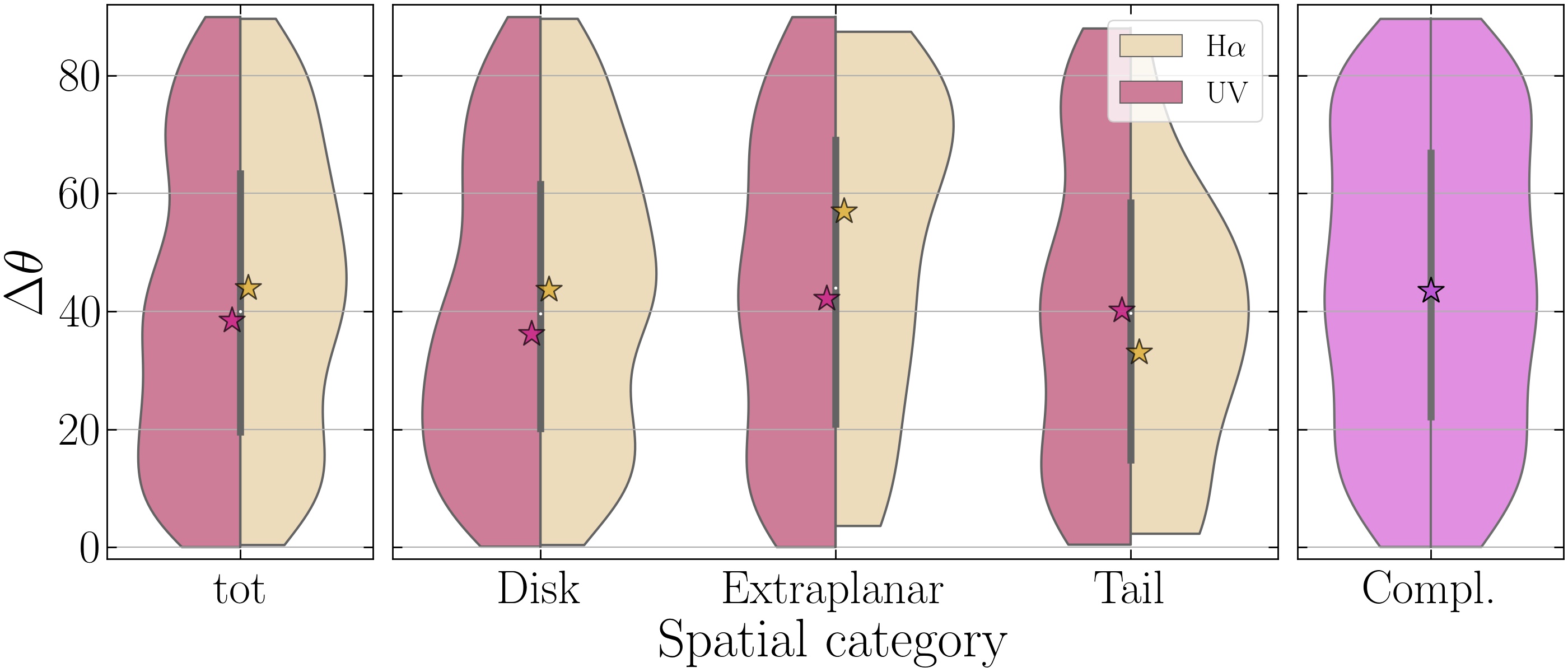}
\caption{Same as Fig. \ref{plot:ar}, but for tilt angle distributions of resolved clumps and complexes.
}
\label{plot:tilt}
\end{figure*}

In Fig. \ref{plot:tilt} we show the violin plots of $\Delta\theta$ for the whole sample of clumps and complexes (left and right panel), and for the clumps divided in spatial category (middle panel). Distributions cover the whole range of possible values of $\Delta\theta$ and are almost flat. Median values are almost consistent with 40-45$^{\circ}$ in the majority of cases, the only exceptions being H$\alpha$-resolved extraplanar and tail clumps, for which we have a tendency for clumps with respectively large and small $\Delta\theta$.

The same distributions are studied for clumps of individual galaxies, at least for the cases in which the number of objects is sufficient to build a distribution. 
No major differences are found with the distributions obtained from the whole samples.

When comparing the projected distances from the center of the galaxy of the H$\alpha$, UV and optical centers of each complex, we select only those cases in which the displacement is larger than our precision on the reciprocal evaluation of the two centers (defined in Sec. \ref{sec:displacement} as the sum in quadrature of the size of two pixels, $0.055$ kpc).
Nonetheless, quantifying the fraction of complexes showing a fireball morphology is not trivial, due to the peculiar morphology of the complexes and the clumps inside them (as it can be seen by the RGB images).
In $66/107$ ($\simeq 62\%$) H$\alpha$-selected matched clumps and $104/187$ ($\simeq 56\%$) UV-selected matched clumps, the clump center is farther than the one of the complex, hinting that the youngest stellar populations have a preferential direction of displacement, far from the galactic disk, as predicted by the fireball model.

It is expected that, under fortunate geometrical alignment, such as ram-pressure stripping occurring on the plane of the sky, the fireball structure is observed, and the brightest H$\alpha$-selected clump is significantly more displaced than the brightest UV-selected clump inside the same complex. We have evidence of that from the RGB images already shown in Fig. \ref{plot:rgb}, with clear evidence of compact H$\alpha$ emission (in red) embedded in diffuse UV emission (in blue), which shows a tail-like structure pointing towards the galaxy.

\section{Searching for trends in the tails}\label{sec:radial}
In principle, the properties of clumps and complexes formed from stripped gas may depend on many factors: when the gas collapsed and how far from the disk, how long ago the clump/complex formed, what was the stage of stripping when the gas collapsed. These factors could influence the properties of a clump as a function of its distance from the galaxy, therefore it is interesting to analyze how the number and properties of clumps and complexes depend on the projected distance from the galaxy center.
In doing this, we consider all galaxies together, after having checked that our trends are not biased by a single galaxy trend and that the same correlations (or absence of correlations) are found also when looking at individual galaxies.

\begin{figure}
\includegraphics[width=0.45\textwidth]{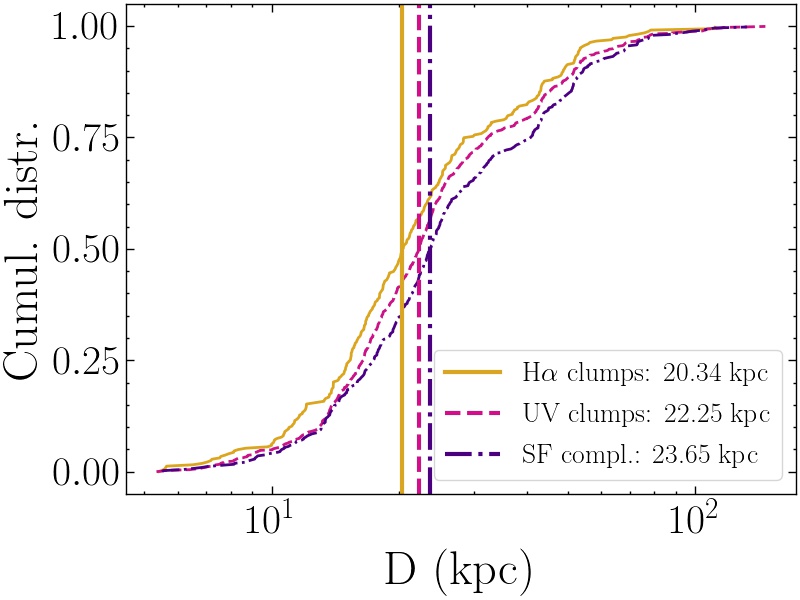}
\caption{Cumulative distribution of tail H$\alpha$- and UV-selected clumps (solid golden and dashed purple, respectively) and star-forming complexes (dash-dotted violet) as a function of the distance. The median distances for each sample of clumps/complexes are plotted as vertical lines with same color and style as the corresponding cumulative distribution.
}
\label{plot:radialhist}
\end{figure}

In Fig. \ref{plot:radialhist} we plot the cumulative distribution of H$\alpha$- and UV-selected clumps and star-forming complexes in the tails as a function of the projected distance D from the hosting galaxy. 
The majority of clumps and complexes lie within 100 kpc from the galaxy center, with a median around 20-24 kpc. The median distance slightly increases from $20.34$ kpc for H$\alpha$-selected clumps, to $22.25$ kpc for UV-selected clumps, to $23.65$ kpc for star-forming complexes
In particular, H$\alpha$ selected clumps are located, on average, closer to their galaxy disks, as shown by the median distances in Fig. \ref{plot:radialhist}, suggesting that most of the star formation occurs soon after the gas is stripped and does not last too long after that.

\begin{figure}
\includegraphics[width=0.48\textwidth]{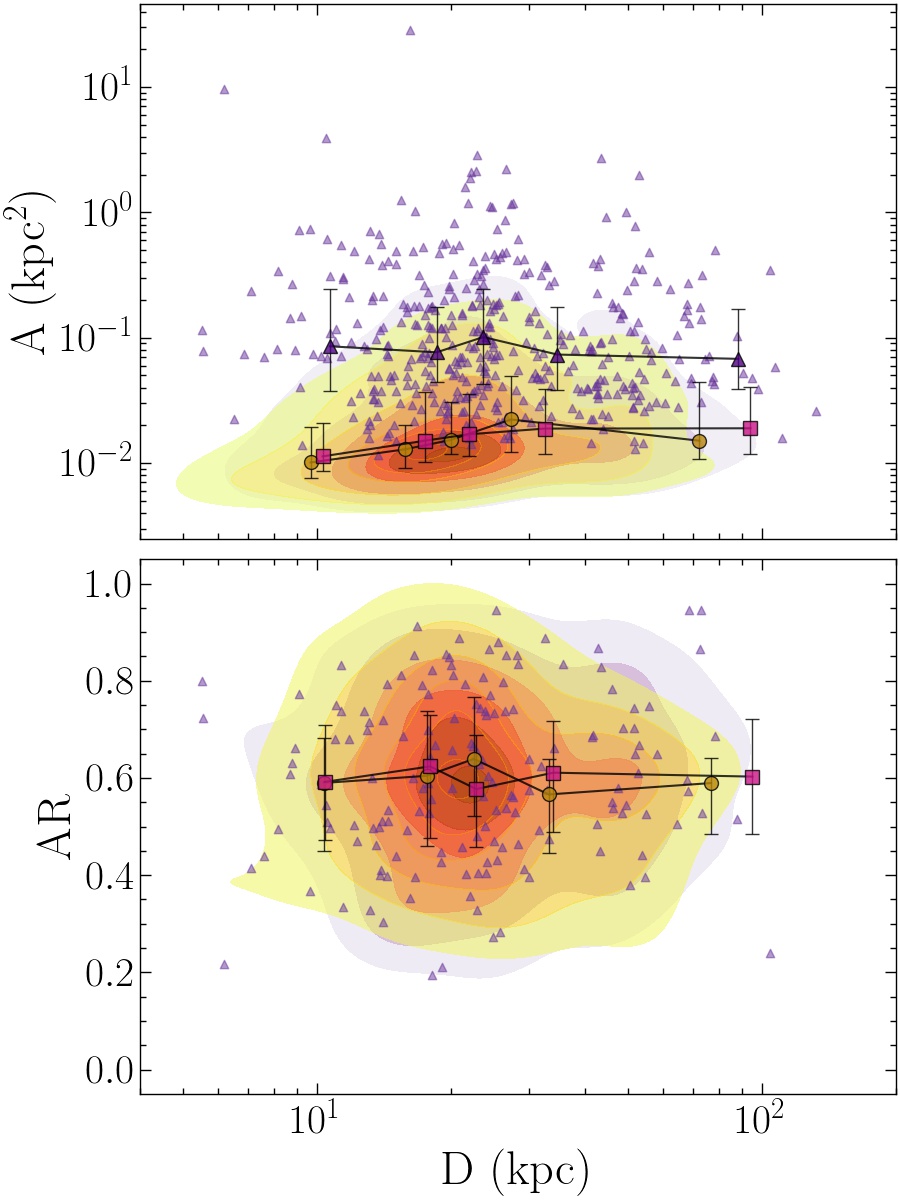}
\caption{\rr{Plots of resolved clump and complex area A (top) and axial ratio AR (bottom) as a function of the projected galactocentric distance.}
Yellow-to-red and magenta contours are the H$\alpha$- and UV-selected/resolved clump density distribution, respectively. Star-forming complexes are plotted as violet triangles. Golden dots, purple squares and violet triangles are the median profiles of the H$\alpha$-selected/resolved clump, UV-selected/resolved clump and star-forming complex properties for bins containing an equal number of clumps/complexes, where the bars indicate the quartiles of the distribution.}
\label{plot:radial}
\end{figure}

We now focus on whether clump and complex characteristics depend on galactocentric distance.
In Fig. \ref{plot:radial} we consider \rr{the clump/complex area and the axial ratio}.
We do not find any evidence of trends for any clump/complex property as a function of distance. The lack of any trend is unclear, but it might be due to a combination of a large number of physical processes influencing gas collapse, star formation and clump distribution that may cancel any possible trend.

\begin{figure}[ht!]
\includegraphics[width=0.48\textwidth]{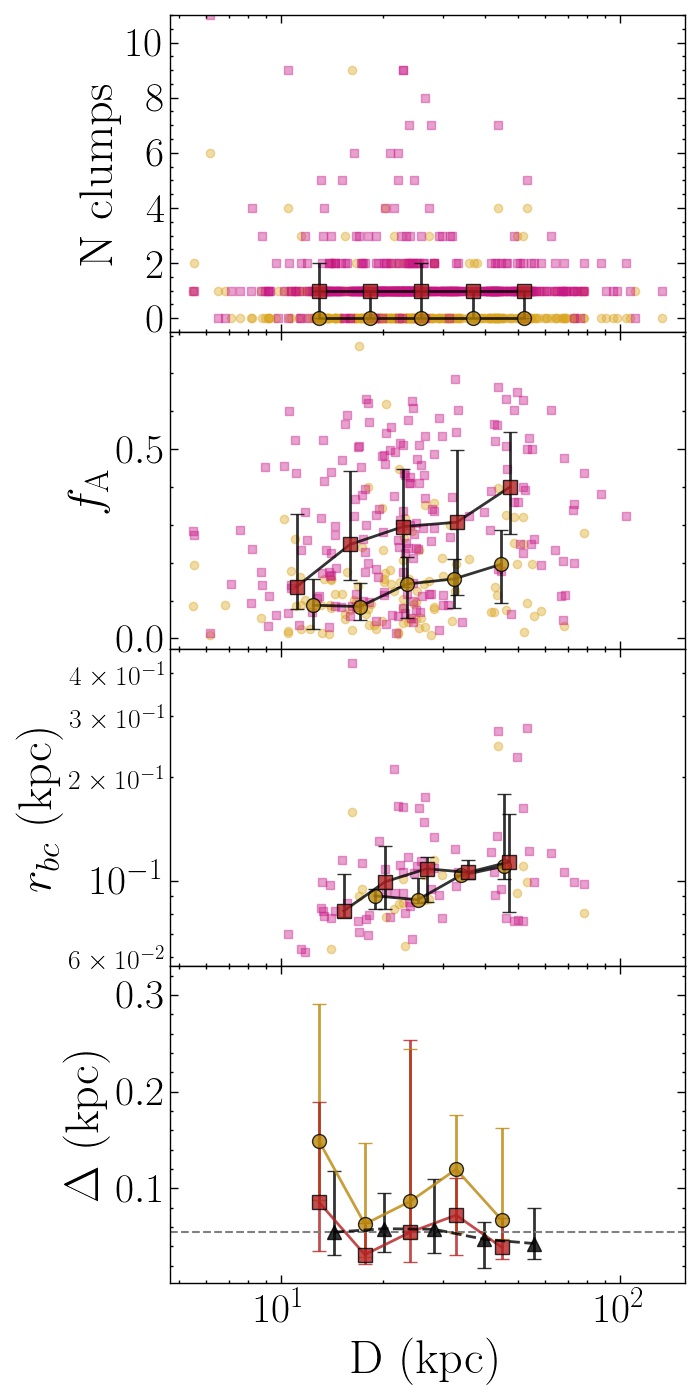}
\caption{Trends with the projected galactocentric distance of the complex properties related to the matched clumps: golden dots when related to matched H$\alpha$-selected clumps, magenta squares to matched UV-selected clumps and black triangles, when present, to UV-selected clumps in complexes with no H$\alpha$-selected clumps. Median profiles are plotted as dark golden dots for H$\alpha$-selected clumps and magenta squares for UV-selected clumps. Error-bars show the quartiles of the distribution in the bin.
From top to bottom: number of matched clumps; filling factor $f_\mathrm{A}$; radius of the brightest matched resolved clump $r_{bc}$; geometric center displacements (defined in Sec. \ref{sec:displacement}, with our resolution limit of $\sim 0.055$ kpc plotted as horizontal dashed grey line).
}
\label{plot:radialmatch}
\end{figure}

We also study the trends with distance of the complex properties which are related to those of the matched clumps.
In Fig. \ref{plot:radialmatch} we plot the radial profiles of four complex quantities that can be derived considering the properties of the H$\alpha$- and UV-selected clumps matched to the complex: number of clumps, $f_\mathrm{A}$, radius of the brightest resolved matched clump, center distances ($\haopt$, $\uvopt$, $\uvoptonly$, definitions in Sec. \ref{sec:displacement}). 
The plots show the presence of two weak positive correlations with $f_\mathrm{A}$ and $r_{bc}$, especially for UV clumps. These trends may be a consequence of the fireball model: since the different generations of stars are characterized by a difference in velocity, the reciprocal distance among them increases with time, with the net result that the clump they form increases in size and better fill the complex to which it is matched.
The same trends are studied computing the distance along their sub-tail (defined as in the appendix \ref{app:subtail}), but no changes are found.

We conclude that the quantities related to the morphology of clumps and complexes and the clumps nesting in the complexes do not correlate with distance, probably as a consequence of the large variety of processes competing in driving and influencing the star formation. In a future work \rr{(Werle et al. in prep.)} we will study the properties of the stellar populations for each clump and complex, like the stellar mass, the star formation rate and the age, by modeling the spectral energy distribution of the 5 filters available.

\section{Conclusions}\label{sec:conclusions}
We have used multi-band \textit{HST} images from \cite{Gullieuszik2023} to analyze the morphological properties of star-forming clumps and complexes of 6 strongly RPS galaxies from the GASP survey \citep{Poggianti2017a}. Star-forming regions are observed in the disk of these galaxies, in the extraplanar regions close to the disk and far from the galaxy (up to 100 kpc) in the stripped tail. The catalogues of H$\alpha$ and UV clumps and of optical complexes were presented in \cite{Giunchi2023}, where we also studied their sizes, luminosity and size distribution functions and luminosity-size relations.

With 2406 H$\alpha$-selected clumps and over 3745 UV-selected clumps spread in the disks, extraplanar regions and tails, and 424 star-forming complexes in the tails, it has been possible to conduct a statistically significant analysis, whose main results can be summarized as follows:

\begin{enumerate}
    \item The axial ratio (i.e. ratio of minor and major axes) of resolved clumps and complexes varies from almost 1 (round objects) to less than $0.2$ (strongly elongated ones). Median values are in the range $0.5-0.7$ in all environments and for all bands, though we see a weak increase in these median values (i.e. rounder clumps) going from the disk to extraplanar to the tail regions. Both very elongated (AR$<0.3$) UV-resolved clumps and optical complexes are preferentially found in the tails. \rr{Most likely, this is a consequence of the RPS acting on the gas from which these objects are formed and the consequent fireball configuration. The fireball can be observed only if the age spread is large enough, which may explain why the elongation is observed only in UV and optical bands, which cover a large range in ages (up to $\sim200$ Myr), and not in H$\alpha$}.

    \item Most complexes host at least one or more (up to 7) UV-selected clumps, while only $37\%$ of them contain at least one H$\alpha$-selected clump, most likely due to ageing effects.
    The number and the size of the clumps correlate with both the area and luminosity of the hosting complex, meaning that large complexes contain more and larger clumps, both in H$\alpha$ and UV.
    
    \item By studying the complexes filling factor, we find that UV-selected clumps better fill the complex than the H$\alpha$-selected clumps. Indeed, the UV median filling factor is $0.27$, compared to the H$\alpha$ median value, $0.1$.
    Both the UV and H$\alpha$ filling factors show a correlation only with the axial ratio of the complex, while they do not correlated with the area of the complex, the number of matched clumps and their sizes.
    Our interpretation is that, at a fixed axial ratio, 
    the number and size of the matched clumps grow with increasing complex area, resulting in a constant filling factor.
    
    \item The relative positions of the $\rm H\alpha$, UV and V-band emission within a complex are not random.
    The average displacement between the center of the V-band emission and the center of the brightest embedded clump increases with the luminosity and the area of the complex, going from a few tens of parsecs up to one kiloparsec. When controlling for these quantities, it is larger for $\rm H\alpha$ than for UV clumps.
    The displacement of complexes with no H$\alpha$-selected clumps is typically in between those of H$\alpha$- and UV-selected for the other properties, with the difference that it is spread over smaller ranges of values (with the only exception of H$\alpha/$UV).
    Interestingly, in the case of the correlation of the displacement with the area and luminosity of the complex, the $\uvoptonly$ seem to follow the same trend as $\haopt$, suggesting that the UV-selected clumps observed in no-H$\alpha$ complexes are former H$\alpha$ clumps, now too old to emit in H$\alpha$ but with the same morphological properties.
    The displacement increases for more elongated complexes, indicating that especially in these ones the clumps tend to occupy one side of the complex. This means that, within a complex, regions of different ages form also a spatial sequence, as confirmed by the fact that the largest displacements are observed for regions with a high $\rm H\alpha$/UV ratio.
    The emerging picture is that clumps are located on one side of the complex, with the $\rm H\alpha$ emission on the extreme side, and show a long optical extension in the other direction.

    \item In many cases, the distribution of clumps within a complex has a preferential direction.
    In many cases, even if the clump elongations have no preferential direction, the clump center is located at larger distance from the galaxy than the complex center, in agreement with the expectations for the "fireball model".
    Not all complexes, however, have a fireball structure, as it is expected given the relevance of projection effects.
    
    \item Clumps and complexes in our sample are observed out to over 100 kpc (projected) from the galaxy center. While UV-selected clumps cover the whole range of distances, H$\alpha$-selected clumps are slightly more concentrated close to the galactic disk.
    Trends with distance of $f_\mathrm{A}$ and $r_{bc}$ are observed. These correlations suggest that the different generations of stars get away from each other as a consequence of the velocity gradient induced by the ram pressure, with the net result that the clump increases in size and better fill the complex in which it is embedded.
    
\end{enumerate}

In this paper we have characterized the morphological properties of clumps within and outside galactic disks, with a specific focus on the nesting properties of clumps and complexes observed in the stripped tails at different wavelengths. In following works, we will present the stellar population properties of these clumps, such as their stellar ages, star formation rates and histories and stellar masses. 
\rr{Combining the results on mass and age with those about size and morphology described in this paper, we will be able to draw solid conclusions about their fate in the galaxy cluster. For instance, the displacement between clumps and complexes can be combined with the age of the different components, putting constraints on the reciprocal velocity of the different stellar generations, which may be a non-negligible quantity to consider when modeling the evolution of the complexes.
The complex may remain bound or not (in the latter case it will disperse), and may fall back onto the parent galaxy and remain bound to it (becoming a local satellite or a stellar stream) or be lost in the galaxy cluster, contributing to the intracluster light either as a small compact object (e.g. a low surface brightness dwarf galaxy, Werle et al. in prep.) or as a diffuse component.}

\begin{acknowledgments}
EG would like to thank \rr{the anonymous referee and} the GASP team for the useful discussions and comments. We would like to thank ESA/Hubble \& NASA for the color-composite images of our sample of galaxies, which are absolutely precious from many points of view.
This paper is based on observations made with the NASA/ESA Hubble Space Telescope obtained from the Space Telescope Science Institute, which is operated by the Association of Universities for Research in Astronomy, Inc., under NASA contract NAS 5-26555. These observations are under the programme GO-16223. All the \textit{HST} data used in this paper can be found in MAST: \dataset[10.17909/tms2-9250]{http://dx.doi.org/10.17909/tms2-9250}.
This paper used also observations collected at the European Organization for Astronomical Research in the Southern Hemisphere associated with the ESO programme 196.B-0578.
This research made use of Astropy, a community developed core Python package for Astronomy by the Astropy Collaboration (\citeyear{Astropy2018}). This project has received funding from the European Research Council (ERC) under the European Union’s Horizon 2020 research and innovation programme (grant agreement No. 833824) and "INAF main-streams" funding programme (PI B. Vulcani).
\end{acknowledgments}

\bibliography{biblio}{}
\bibliographystyle{aasjournal}

\appendix

\section{Sub-tails definition}\label{app:subtail}
\begin{figure*}
\centering
\gridline{\resizebox{\textwidth}{!}{\includegraphics[height=1cm]{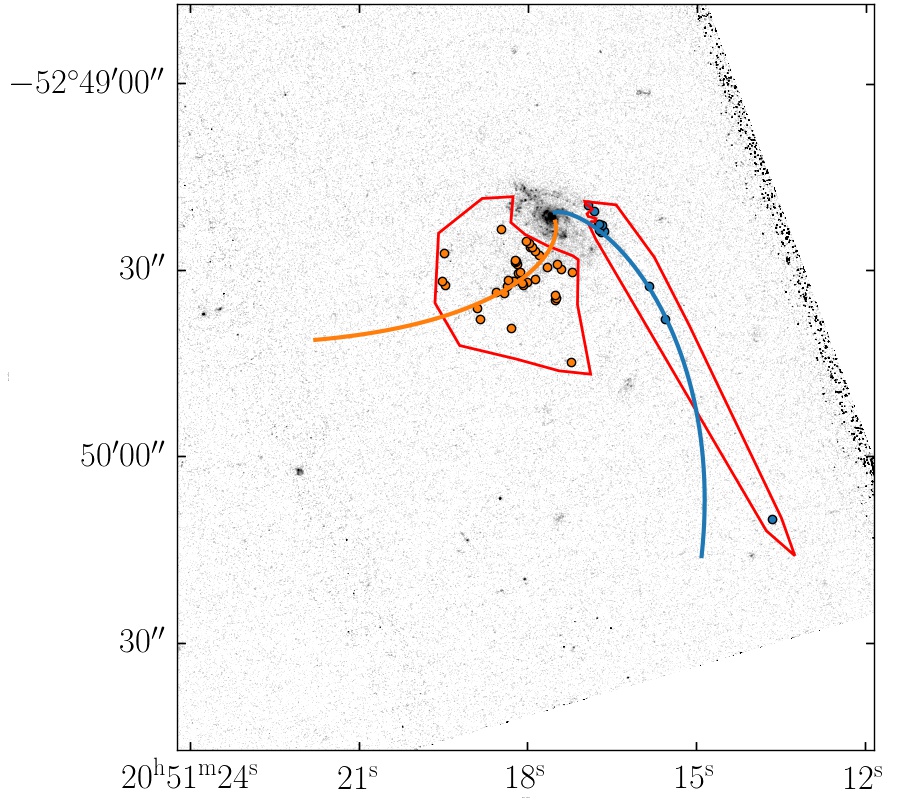}
          \includegraphics[height=1cm]{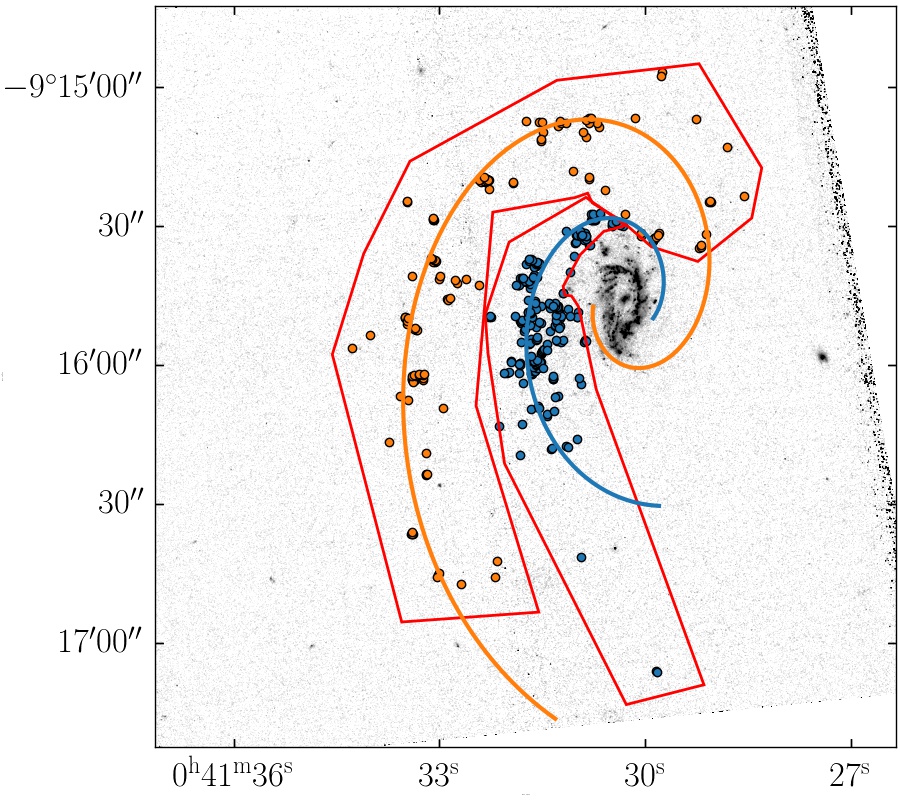}
          \includegraphics[height=1cm]{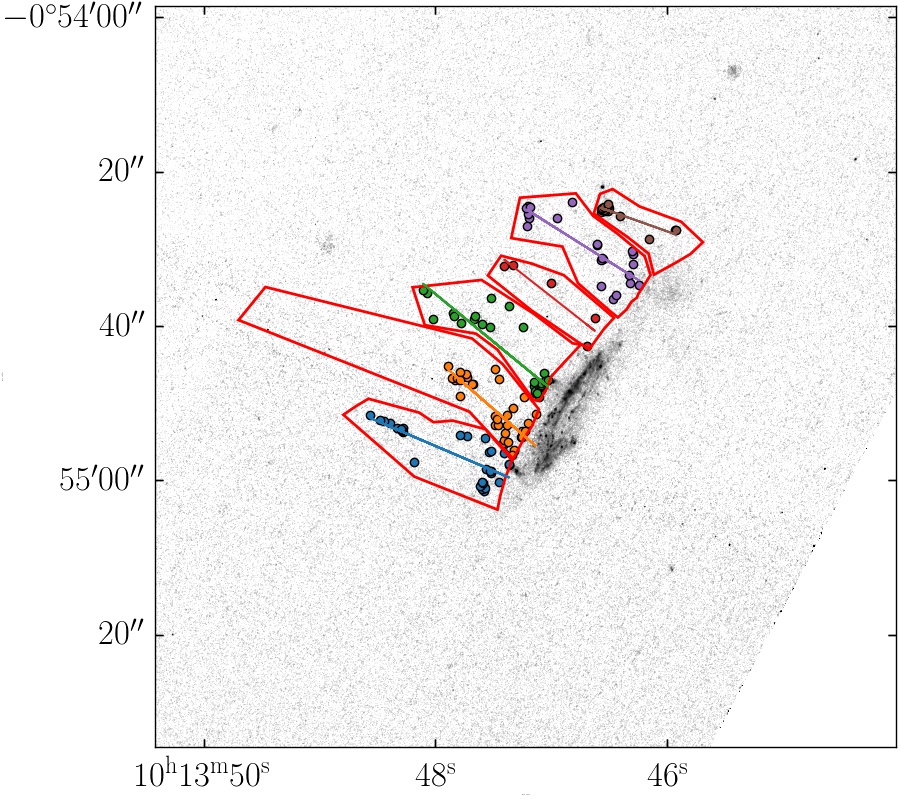}
          }
          }
\gridline{\resizebox{\textwidth}{!}{\includegraphics[height=1cm]{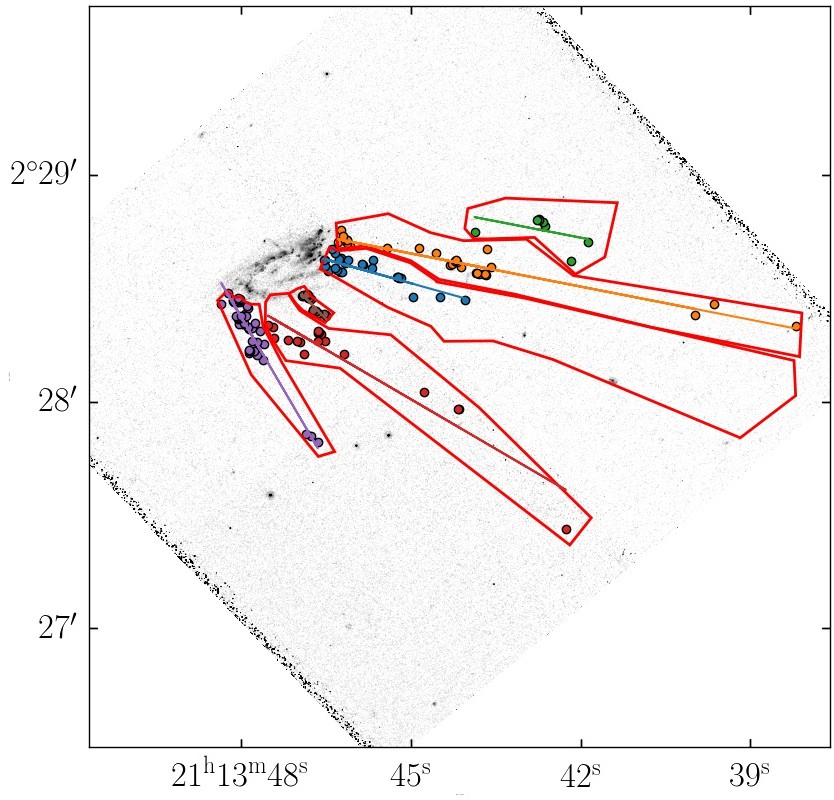}
          \includegraphics[height=1cm]{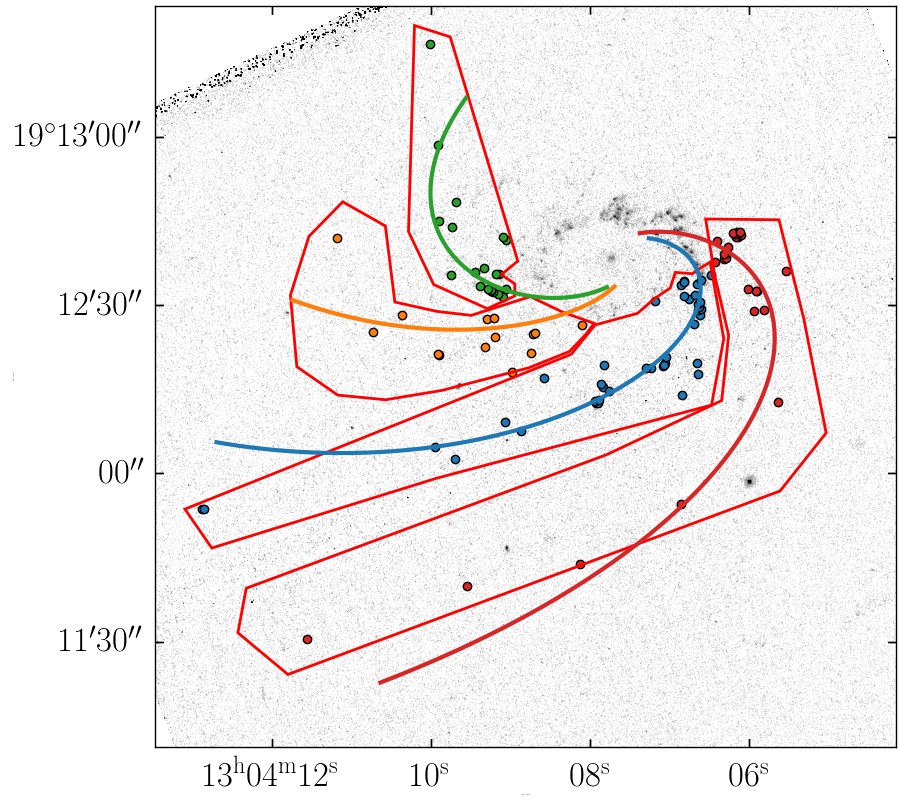}
          \includegraphics[height=1cm]{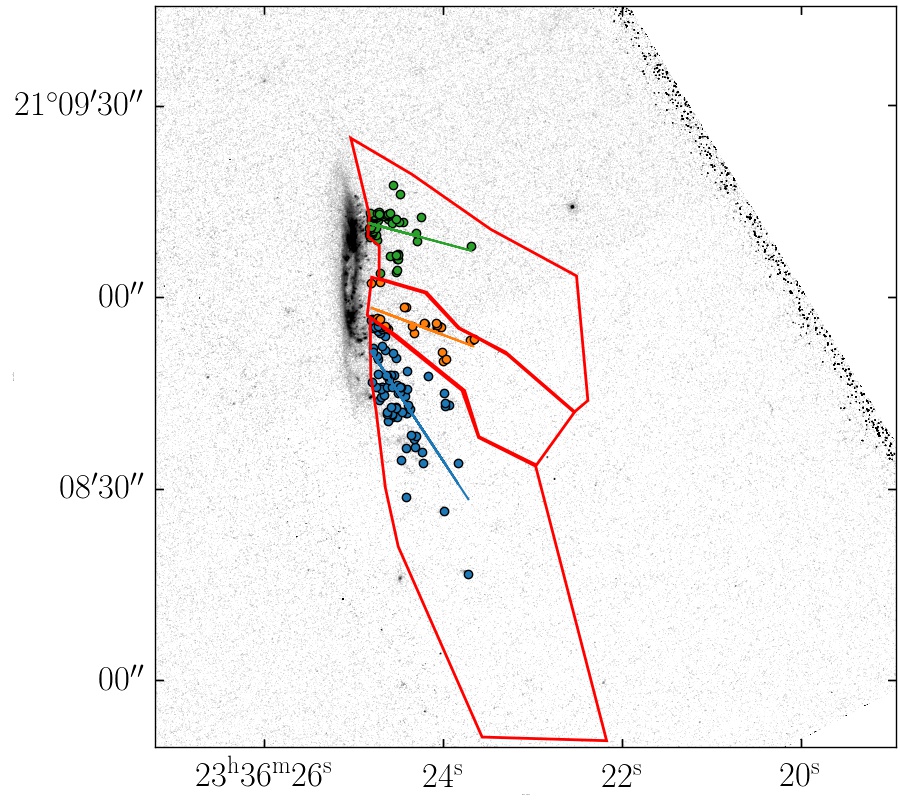}
          }}
\caption{F275W maps of the six galaxies of our sample. Top row (from left to right): JO175, JO201, JO204; bottom row: (from left to right): JO206, JW39, JW100. In different colors we plot the sub-tails we defined for each galaxy: the dots are the UV-selected clumps we used to trace the regions (solid, closed regions) defining the sub-tails; the best-fitting trajectories of the sub-tails are plotted as solid lines (as straight lines or spiral arms, according to the choice made in Sec. \ref{app:subtail}). The red dots are the zero-points of the sub-tails, from which the distances of the clumps along the sub-tails are computed. Each sub-tail is flagged with a number.
}
\label{plot:subtails}
\end{figure*}

Since we are interested in studying the distribution of the properties of the clumps along the tails, for each galaxy the UV-selected clumps aligned to each other forming a chain and with a similar gas kinematics are grouped in sub-tails (Fig. \ref{plot:subtails}).
MUSE maps of the gas kinematics are presented and discussed in \cite{Poggianti2017b}, \cite{Poggianti2017a}, \cite{Bellhouse2017} and \cite{Gullieuszik2017}. To properly estimate the projected distance of a clump from the galaxy along each sub-tail, we defined the sub-tails according to the following two definitions, according to the stripping morphology:
\begin{enumerate}
    \item{linear: as done in \cite{Franchetto2021}, the positions of the clumps in the plane of the sky are fitted with a line. The distances of the clumps are computed projecting the clump positions onto the best-fitting line and fixing the zero-point to the projection of the first clump of the sub-tail. This definition is adopted for JO204, JO206 and JW100 (i. e. the nearly edge-on galaxies);}
    \item{logarithmic spiral arm-like: in this case, we followed the same procedure described in \cite{Bellhouse2021}. Each galaxy F275W image is deprojected according to the axial ratios measured in \cite{Franchetto2020}, by scaling the distances along the dimension of the kinematic minor axis. 
    The position in the plane of the sky of each pixel is then converted in spherical polar coordinates according to its radial distance from the centre of the galaxy and azimuthal position. In this space (in particular in the $\log r-\theta$ plane), structures with a logarithmic spiral shape appear as straight lines, and therefore we manually fit a line to the tails of the galaxies\footnote{We point out that the deprojection cannot take into account the vertical distance of the clumps in the tails from the plane of the disk, as the distance among the tail clumps and between the clumps and the galaxy cannot be recovered.}. As explained in \cite{Bellhouse2021}, we opted for a manual method because the tails of these galaxies are characterized by very disturbed and peculiar morphologies, which are not easily fitted using an automated process. When possible, we used the same curves fitted in \cite{Bellhouse2021}, adding new ones where necessary. 
    Finally, for each sub-tail, clumps are projected onto their spiral arm, and their distances are computed from these points to the zero-point, defined as the projection of the first clump of the sub-tail. This definition is adopted for JO175, JO201 and JW39 (i. e. the nearly face-on galaxies).}
\end{enumerate}

The distances are then converted in kpc according to the redshift of the galaxy cluster.

\end{document}